\documentclass[aps,prd,preprint,eqsecnum,nofootinbib,groupedaddress,showpacs,floatfix]{revtex4}
\usepackage[dvips]{graphicx}
\font\ttiny=cmr6 at 3.5pt

\newcommand{\BlackHat}{{\sc BlackHat}}
\newcommand{\SHERPA}{{\sc SHERPA}}
\newcommand{\AMEGIC}{{\sc AMEGIC++\/}}
\newcommand{\SISCone}{{\sc SISCone}}

\newif\ifdraft
\drafttrue
\newif\ifpreprint
\preprinttrue

\def\be{\begin{equation}}
\def\ee{\end{equation}}
\def\bea{\begin{eqnarray}}
\def\eea{\end{eqnarray}}
\def\fig#1{fig.~{\ref{#1}}}
\def\Fig#1{Fig.~{\ref{#1}}}
\def\figs#1#2{figs.~{\ref{#1}} and {\ref{#2}}}

\def\sect#1{section~{\ref{#1}}}
\def\eqn#1{eq.~(\ref{#1})}

\def\eqns#1#2{eqs.~(\ref{#1}) and~(\ref{#2})}

\def\tab#1{table~{\ref{#1}}}
\def\Tab#1{Table~{\ref{#1}}}

\def\ths{{\theta^*}}
\def\thstilde{{\tilde\theta^*}}
\def\phis{{\phi^*}}

\def\pol{\varepsilon}
\def\tree{{\rm tree}}
\def\spa#1.#2{\left\langle#1\,#2\right\rangle}
\def\spb#1.#2{\left[#1\,#2\right]}

\def\Wj{$W\,\!+\,1$}
\def\Wjj{$W\,\!+\,2$}
\def\Wjjj{$W\,\!+\,3$}

\def\Wjjja{$W\,\!+\,1,2,3$}
\def\Wjjjja{$W\,\!+\,1,2,3,4$}
\def\Wjjjx{$W\,\!+\,3,4$}
\def\Wjn{$W\,\!+\,n$}

\def\Wpj{$W^+\,\!+\,1$}
\def\Wpjj{$W^+\,\!+\,2$}
\def\Wpjjj{$W^+\,\!+\,3$}

\def\Wpjjja{$W^+\,\!+\,1,2,3$}

\def\Zj{$Z\,\!+\,1$}

\def\Zgamjjj{$Z,\gamma^*\,\!+\,3$}

\def\nn{\nonumber}

\def\HTpartonicp{{\hat H}_T'}

\newbox\charbox
\newbox\slabox
\def\s#1{{      
        \setbox\charbox=\hbox{$#1$}
        \setbox\slabox=\hbox{$/$}
        \dimen\charbox=\ht\slabox
        \advance\dimen\charbox by -\dp\slabox
        \advance\dimen\charbox by -\ht\charbox
        \advance\dimen\charbox by \dp\charbox
        \divide\dimen\charbox by 2
        \raise-\dimen\charbox\hbox to \wd\charbox{\hss/\hss}
        \llap{$#1$}
}}

\def\nuornub{{}^{\raise1.3pt\hbox{\ttiny(}}\hskip -0.2pt\overline{\kern 
-0.5pt \nu \kern -0.4pt}\hskip 0.3pt{}^{\raise1.3pt\hbox{\ttiny)}}}

\begin{document}

\hbox{UCLA/11/TEP/105 $\null\hskip 1.8cm \null$
CERN--PH--TH/2011-062 $\null\hskip 1.8cm \null$
SLAC--PUB--14409}
\hbox{
SB/F/386-11 $\null\hskip 1.0cm \null$
NIKHEF-2011-006 $\null\hskip 1.0cm \null$
Saclay--IPhT--T11/040 $\null\hskip 1.0cm \null$
IPPP/11/15}

\vskip.3cm

\title{Left-handed $W$ bosons at the LHC}
\author{Z.~Bern${}^a$,
 G.~Diana${}^b$, L.~J.~Dixon${}^{c,d}$,
F.~Febres Cordero${}^e$,
 D.~Forde${}^{c,f}$,
T.~Gleisberg${}^d$, S.~H{\"o}che${}^d$, H.~Ita${}^a$,
D.~A.~Kosower${}^b$,
D.~Ma\^{\i}tre${}^{c,g}$ and K.~Ozeren${}^a$
\\
$\null$
\\
${}^a$Department of Physics and Astronomy, UCLA, Los Angeles, CA
90095-1547, USA \\
${}^b$Institut de Physique Th\'eorique, CEA--Saclay,
          F--91191 Gif-sur-Yvette cedex, France\\
${}^c$Theory Division, Physics Department, CERN, CH--1211 Geneva 23, 
    Switzerland\\
${}^d$SLAC National Accelerator Laboratory, Stanford University,
             Stanford, CA 94309, USA\\
${}^e$ Universidad Sim\'on Bol\'{\i}var, Departamento de
F\'{\i}sica, Caracas 1080A, Venezuela\\
${}^f$NIKHEF Theory Group, Science Park 105, NL--1098~XG
  Amsterdam, The Netherlands\\
${}^g$Department of Physics, University of Durham,
          DH1 3LE, UK\\
}


\begin{abstract}
The production of $W$ bosons in association with jets is an important
background to new physics at the LHC.  Events in which the $W$ carries
large transverse momentum and decays leptonically lead to large
missing energy and are of particular importance.  We show that the
left-handed nature of the $W$ coupling, combined with valence quark
domination at a $pp$ machine, leads to a large left-handed
polarization for both $W^+$ and $W^-$ bosons at large transverse
momenta.  The polarization fractions are very stable with respect to
QCD corrections.  The leptonic decay of the $W^\pm$ bosons translates
the common left-handed polarization into a strong asymmetry in
transverse momentum distributions between positrons and electrons, and
between neutrinos and anti-neutrinos (missing transverse energy).
Such asymmetries may provide an effective experimental handle on
separating $W$\,+\,jets from top quark production, which exhibits very
little asymmetry due to C invariance, and from various types of new
physics.
\end{abstract}


\maketitle


\section{Introduction}
\label{IntroductionSection}

Events produced by new physics at the LHC often resemble events
generated by Standard Model physics.  This is especially true for
signals involving multiple jets alongside a $W$ boson that decays 
to a lepton pair.  These kinds of events most commonly arise from
QCD emission in
an electroweak process.  Such events could also be the result of cascade
decays in supersymmetric extensions of the Standard Model, as
well as in other models of physics beyond the Standard Model. They
also emerge in top-quark pair production in its semileptonic decay mode,
and in some Higgs search modes.
The QCD $W\,\!+$ jets events pose a background to all of these signals.

The superficial similarity of such signals to overwhelmingly larger
backgrounds pushes us to find differences in various
distributions, so as to impose cuts that suppress the Standard-Model
backgrounds while retaining as much of the signals of new physics as possible.
General underlying properties or principles that distinguish different sources
of similar events are of particular importance.  In this paper, we discuss
one such general property, the large left-handed polarization of
high-$p_T$ ``prompt'' $W^+$ and $W^-$ vector
bosons produced directly in short-distance Standard-Model 
interactions~\cite{W3jDistributions}.
This effect is distinct from the well-known~\cite{ESW}
left-handed polarization of
low-$p_T$ $W$ bosons moving primarily along the beam axis.

The importance of \Wjn-jet final states in hadron-collider searches
has prompted intensive theoretical work over the last two decades.
Leading-order (LO) matrix-element generators have been available
for some time~\cite{LOPrograms,Amegic}.  More recently, they have
been combined with parton-shower approaches using several matching
techniques~\cite{Matching}, in order to provide event simulations
which combine the correct wide-angle properties (at LO) with the detailed
intrajet particle distributions required by experimenters.  LO predictions,
however, leave the overall normalization of event rates uncertain,
an uncertainty that rises as the number of jets increases.

Obtaining quantitatively reliable predictions for \Wjn-jet rates and
distributions requires next-to-leading order (NLO) cross sections in
QCD.  The one-loop matrix elements entering NLO predictions stabilize
the dependence on the unphysical renormalization and factorization
scales, and provide predictions expected to be reliable to 10--15\%.
Until recently, calculating the required one-loop matrix elements
posed a major difficulty, especially for final states with many jets
(and hence many partons).  Unitarity-based techniques have broken 
this bottleneck, and have allowed the prediction at NLO of vector-boson
production with up to four associated
jets~\cite{BHPRL,W3jDistributions,Ellis3j,BHZ3j,BHW4j}.  These
predictions do indeed possess greatly reduced overall normalization
uncertainties.  They also indicate where the LO predictions for shapes
of distributions are reliable, and where they suffer corrections.  For
more than one associated jet, the results are available at the parton
level, and have not yet been incorporated into a matched parton-shower
maintaining NLO accuracy.

\begin{figure*}[tbh]
\begin{minipage}[b]{1.\linewidth}
\includegraphics[clip,scale=.4]{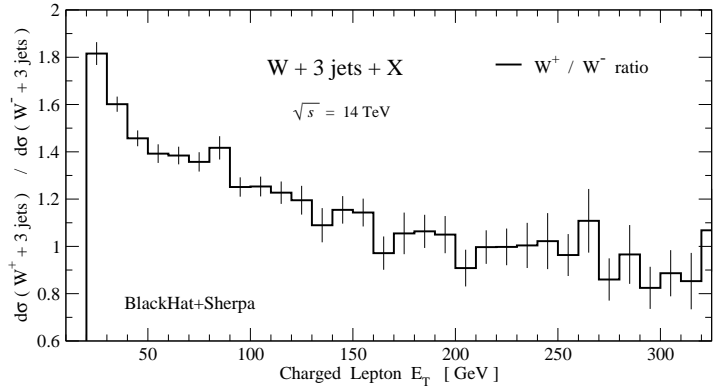}
\end{minipage}
\caption{ The ratio of the charged-lepton $E_T$ distributions at the
LHC for $W^+$ and $W^-$ production in association with three jets,
evaluated at NLO~\cite{W3jDistributions}.}
\label{ChargedRatioFigure}
\end{figure*}

It was first observed in ref.~\cite{W3jDistributions} that for
\Wjjja-jet events at the LHC, both $W^+$ and $W^-$ bosons produced via
Standard-Model interactions are preferentially polarized left-handed
along their flight direction, beyond small transverse momenta.  The
polarization manifests itself in the decay spectra of the daughter
leptons: left-handed $W^+$ bosons at a fixed boson $p_T^W$ produce
larger neutrino transverse momentum (missing $E_T$) and smaller
charged lepton $p_T$, in comparison with the decays of left-handed
$W^-$ bosons.  The polarization thus gives a characteristic shape to
the ratios of the charged-lepton $E_T$ spectra, as shown in
\fig{ChargedRatioFigure}, between $W^+$ and $W^-$ production in
association with three jets.  (The precise setup used for this plot
may be found in ref.~\cite{W3jDistributions}; the key feature, a falling 
ratio with increasing $E_T$, is generic.)  
An opposite but similarly characteristic
shape arises in the ratio of missing $E_T$ distributions for $W^+$
versus $W^-$ events.  These ratio distributions are quite stable upon
going from LO to NLO.  Ref.~\cite{Polarization} computed directly the
left, right and longitudinal polarization fractions $f_L$, $f_R$ and
$f_0$ for the case of \Wpjj-jet production at the LHC.  The
polarization fractions are quite stable over a range of $W$ transverse
momenta, with $f_L$ of order 60\% and rising slowly with $p_T^W$,
$f_R$ of order 25\%, and the remaining longitudinal fraction $f_0$
dropping monotonically toward zero as $p_T^W$ increases.  The
vanishing of $f_0$ at large $p_T^W$ is dictated by the equivalence
theorem~\cite{Equivalence}, because the Goldstone modes cannot couple
to light quark lines.

In this paper we explore the dynamics behind the production of left-handed
prompt $W$ bosons at finite transverse momentum at the LHC.
We explain the underlying mechanism in terms of a
combination of the left-handed nature of the charged-current
weak interactions (which allows only left-handed quarks to participate
at lowest order), the domination of quarks over anti-quarks in the
incoming protons, and the structure of the relevant helicity amplitudes.
  
Our earlier results~\cite{W3jDistributions,Polarization} have prompted
CMS to undertake a measurement of $W$ boson polarization using the
2010 LHC data collected at $\sqrt{s}=7$~TeV~\cite{CMSMeasurement}.
Here we provide theoretical predictions relevant to
this measurement, which require a minimum $p_T^W$ of 50~GeV,
but no explicit requirement of additional jets.  We apply no explicit
cuts on the lepton transverse momenta or rapidity; and no lepton
isolation cuts either.  The experiments do, of course, make such cuts
--- the detectors have finite size and cracks, and triggers impose
implicit lepton $p_T$ cuts.  CMS has corrected its measured data for
the effects of the lepton cuts.
The actual cuts used, combined with the lack of knowledge of the longitudinal
component of the neutrino momentum, lead to a dependence of the
extracted fractions $f_L$, $f_R$ and $f_0$ on other components of the
full $W$ boson spin density matrix.

The diagonal elements of the density matrix (essentially $f_L$, $f_R$
and $f_0$) are coefficients of functions that depend only on the polar
angle $\ths$ of the charged lepton in the $W$ rest frame, with respect
to the $W$ flight direction as observed in the lab frame. The
neutrino will of course come out at an angle $\pi - \ths$ in 
this frame. The
off-diagonal elements arise from the interference of amplitudes for
different $W$ helicity states, and they depend on an azimuthal angle
$\phis$.  They would integrate to zero if the experimental acceptance
were uniform in $\phis$, but it is not.  Accordingly, theoretical
information about the $\phis$ dependence, and its uncertainty, is
needed in order to extract $f_L$, $f_R$ and $f_0$.  The full $W$ spin
density matrix has been studied previously in several theoretical
papers~\cite{CollinsSoper,LamTung,Hagiwara1,Mirkes,MirkesOhnemus,Hagiwara2}.
Two of the additional
coefficients were measured by CDF during Run I of the
Tevatron~\cite{CDFRunI}.

In this paper we compute the diagonal and off-diagonal
elements of the density matrix as a function of the $W$ boson transverse
momentum at the
LHC at $\sqrt{s} = 7$~TeV, expressed as asymmetry coefficients $A_i$
of various angular distributions.  We perform the computation at
LO and at NLO, {\it i.e.}~fixed-order and parton level, using
\BlackHat~\cite{BlackHatI} in conjunction with \SHERPA~\cite{Sherpa}.
We also use \SHERPA\ to provide a parton-shower prediction matched to
tree-level matrix elements, also known as
matrix-element-plus-truncated-shower (ME+PS).  (Our parton-shower
results do not include hadronization effects, but remain at the
parton level.)  We find that the
corrections from LO to NLO are fairly small.  We also find that
varying the factorization and renormalization scale in a correlated
way in the numerator and denominator of the ratios entering the $A_i$
gives very small changes, so small that it does not provide a sensible
measure of the theoretical uncertainty.  We have studied the
dependence of the $A_i$ on the parton distributions, using the error
sets provided by CTEQ~\cite{CTEQ6M}, and find it to be small as well.

We ascribe the principal theoretical uncertainty to the difference
between the NLO and ME+PS results. This difference is typically of
order 10\% for the larger $A_i$ coefficients, including $f_L$, $f_R$
and $f_0$. (The uncertainties from the choice of parton distributions
are significantly smaller.)  Some of the $A_i$ coefficients are quite
small in magnitude, presumably due to cancellations between different
types of terms.  In this case the percentage difference between NLO
and ME+PS can be significantly larger.

This article is organized as follows.  In \sect{DynamicsSection}
we give arguments why $W$ bosons are predominantly left-handed at the LHC.
In \sect{MethodsSection} we define the polarization more 
precisely and explain how we compute it.  Our results are presented
in \sect{ResultsSection}. In \sect{ConclusionSection} we give our
conclusions.


\section{Dynamics of $W$ Polarization at the LHC}
\label{DynamicsSection}

In this section we will explain why both $W^+$ and $W^-$ bosons
produced at the LHC are dominantly polarized left-handed when they emerge
with large transverse momentum.  We will start with a heuristic explanation
based on angular momentum conservation, and then proceed to refine
the explanation further.  A fully quantitative description
requires a numerical calculation, which we present in \sect{ResultsSection}.

\begin{figure*}[tbh]
\begin{minipage}[b]{1.\linewidth}
\includegraphics[clip,scale=.7]{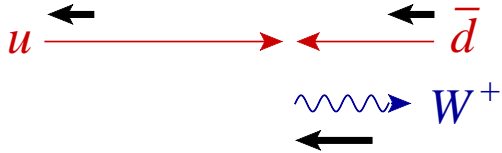}
\end{minipage}
\caption{When a $W^+$ is produced at lowest order by
$u(x_1)\bar{d}(x_2) \to W^+$ with $x_1 > x_2$, it is 100\%
left-handed polarized along its direction of motion, which is along
the beam axis in the quark direction. Thick (black) arrows represent
spin vectors; the other arrows represent momentum vectors in the
$pp$ center-of-mass frame.}
\label{WbeamlineFigure}
\end{figure*}

Before considering the case of $W$ production with transverse momentum,
we discuss the simpler and well-known example of $W$ polarization
along the beam axis, for $W$ bosons produced with little or
no transverse momentum~\cite{ESW}.
Here the principal production mechanism involves the 
leading-order partonic subprocesses $u\bar{d} \to W^+$ and
$d\bar{u} \to W^-$.
At leading order, the $W$ moves strictly along the beam axis, with no
transverse momentum, $p_T^W=0$.  Suppose the $W$ is moving in the 
direction of the
initial-state quark, as opposed to the anti-quark.  This is likely to
be the case at the LHC, because the LHC is a $pp$ machine and the quark
distributions $q(x)$ have a larger average momentum fraction $x$
than the antiquark distributions $\bar{q}(x)$.  Because the electroweak
charged current is purely-left-handed, the quark must be left-handed
and the anti-quark right-handed.  (We assume massless quarks
and leptons throughout this paper.)  By angular momentum conservation,
the spin of the $W$ is 100\% left-handed along its direction of motion,
for either $W^+$ or $W^-$, as shown in \fig{WbeamlineFigure}.
This effect is diluted some by anti-quarks that occasionally carry
a larger $x$ than the quarks with which they collide.  However, the
dilution is small at large rapidities, because the ratio
$q(x)/\bar{q}(x)$ increases rapidly as $x\to1$.

The $W$ polarization is analyzed with 100\% analyzing power through
its leptonic decay.  A left-handed $W^+$ tends to decay with the
left-handed neutrino forward (along its direction of motion) and
the right-handed positron backward.  A left-handed $W^-$ tends to
put the left-handed electron forward and the right-handed anti-neutrino
backward.

Note that at the Tevatron, a $p\bar{p}$ collider, the same basic physics
of valence quark domination and angular momentum conservation
causes the $W^+$ bosons, which typically move in the proton direction,
to be primarily left-handed.  However, the $W^-$ bosons, which typically
move in the anti-proton direction, usually arise from a right-handed $\bar{u}$
anti-quark from the anti-proton annihilating with a left-handed $d$ quark
from the proton; hence the $W^-$ bosons are predominantly right-handed
at the Tevatron.  This polarization implies that both $W^+$ and $W^-$
bosons tend to decay so that the charged leptons are more central
than the parent bosons, producing the well-known dilution of the $W$
boson charge asymmetry, when it is measured via the charged-lepton 
rapidity distribution.

Next consider the case in which the $W$ boson does carry transverse momentum.
For definiteness, we take the $W$ to be a $W^+$; the case of a $W^-$
is qualitatively the same. At leading order, there are three possible
subprocesses:  $ug\to W^+d$, $u\bar{d} \to W^+g$, and $g\bar{d}\to W^+\bar{u}$.
These subprocesses are all related to each other by crossing symmetry.
For events with a sufficiently large $W$ transverse momentum,
the soft-gluon enhancement of $u\bar{d} \to W^+g$ is not that important,
and the hierarchy of subprocess contributions is set by the hierarchy
of relevant parton distributions for typical values of $x$.  Now, $x$
increases with $p_T^W$; at sufficiently large
$x$, $q(x) \gg g(x) \gg \bar{q}(x)$, which leads to the subprocess
hierarchy,
\be
\frac{d\sigma(ug\to W^+d)}{dp_T^W}
\ >\ \frac{d\sigma(u\bar{d} \to W^+g)}{dp_T^W}
\ >\ \frac{d\sigma(g\bar{d}\to W^+\bar{u})}{dp_T^W} \,,
\label{subprocesshierarchy}
\ee
once we include the convolution with parton distributions in the quantities
in \eqn{subprocesshierarchy}.
Even at more moderate $p_T^W$ (smaller $x$), where the second hierarchy might
be small, or even reversed, the first one should still hold.
(Kom and Stirling~\cite{KomStirling} have found that at LO the fraction
of subprocesses in \Wjjjja-jet production that
are initiated by the $qg$ channel (plus $\bar{q}g$) is around 70--80\%.)

\begin{figure*}[tbh]
\begin{minipage}[b]{1.\linewidth}
\includegraphics[clip,scale=.7]{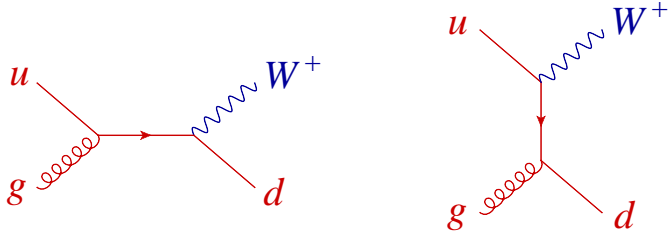}
\end{minipage}
\caption{The two Feynman graphs for the subprocess $ug\to W^+d$.}
\label{ugWdFeynmanFigure}
\end{figure*}

Consider the $W$ polarization produced by the dominant subprocess,
$ug\to W^+d$.  The analysis is more complicated than in the case of
production along the beam axis, because two different axes are
involved, the beamline and the $W$ flight direction.  In this section
only, to simplify the analysis we define the $W$ flight direction
using the partonic center-of-mass frame.  (In subsequent sections we
will use the lab frame; at very high $p_T^W$ there is not much difference
between these choices.)  There are two Feynman graphs for this
process, shown in \fig{ugWdFeynmanFigure}, the $s$-channel graph on
the left and the $t$-channel graph on the right.  We first give an
heuristic argument that the $W$ is left-handed, based on angular
momentum conservation along the $W$ flight direction.\footnote{We
thank Jeff Richman for suggesting this argument.}  Suppose for a
moment that we could neglect the $t$-channel graph.  Then the
subprocess would involve an off-shell spin-1/2 $u$-quark, which decays
to an on-shell, left-handed $d$ quark recoiling against the $W$ boson.
In this case it is impossible for the $W$ boson to be right-handed,
because the total angular momentum along the $W$--$d$ axis would then be
$1+1/2 = 3/2$, which cannot be carried by the spin-1/2 off-shell
quark.  Also, the longitudinal mode of the $W$ is suppressed for large
transverse momenta, $p_T^W \gg M_W$, by the equivalence theorem which
relates this mode to the Goldstone boson, which does not couple to
massless fermions.  Thus we could argue that the $W$ boson is 100\%
left-handed at large $p_T^W$ if only we could neglect the $t$-channel
graph.

\begin{figure*}[tbh]
\begin{minipage}[b]{1.\linewidth}
\includegraphics[clip,scale=.7]{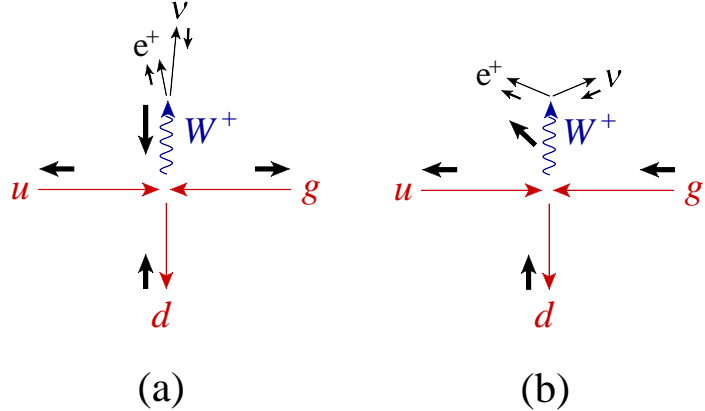}
\end{minipage}
\caption{Helicity configurations for the subprocess $ug\to Wd$ for
(a) a left-handed incoming gluon and (b) a right-handed one.
Thick arrows again denote spin vectors.  In case (a) the
$W$ is purely left-handed, after boosting from the partonic center-of-mass
frame. In case (b) it has indefinite polarization, but becomes purely
right-handed at large $W$ transverse momentum, as indicated by the arrow
at an angle. However, the squared matrix element is smaller than in case (a).}
\label{ugWdFigure}
\end{figure*}

In fact, this argument is true when the incoming gluon is left-handed.
To see this,
we choose the gluon polarization vector so that the $t$-channel graph in
\fig{ugWdFeynmanFigure} vanishes.  This graph contains a factor of
$\s{\pol}^\pm(k_g,q) |k_d^+\rangle$, where $|k_d^+\rangle$ is a
Weyl spinor for the outgoing $d$ quark momentum $k_d$, and 
$\pol_\mu^\pm(k_g,q)$ is the polarization vector for the gluon with
momentum $k_g$.  We use a spinor-helicity representation for this 
polarization vector, in terms of a reference momentum/spinor $q$.
Contracting with $\gamma^\mu$ yields the gluon polarization
bi-spinors,
\bea
\s{\pol}^+(k_g,q) &=&
\frac{ \sqrt{2} \, | k_g^- \rangle \, \langle q^- | }{\spa{q}.{k_g} } \,,
\label{slashpolglueRout}\\
\s{\pol}^-(k_g,q) &=&
- \frac{ \sqrt{2} \, | q^- \rangle \, \langle k_g^- | }{\spb{q}.{k_g} } \,,
\label{slashpolglueLout}
\eea
using a standard all-outgoing labeling of the gluon helicity $\pm$,
where we dropped terms that vanish when contracted with a left-handed
$d$ or $u$ spinor.
For a left-handed incoming gluon, as shown in \fig{ugWdFigure}(a),
we are instructed to use \eqn{slashpolglueRout}.  In this case the
$t$-channel graph is proportional to
$\langle q^- | k_d^+\rangle \equiv \spa{q}.{k_d}$.
We are now free to choose the reference spinor $q=k_d$, so that
$\spa{q}.{k_d}=0$ and thus the $t$-channel graph vanishes.
Although this is a specific gauge choice, it allows us to argue that
the $W$ should be 100\% left-handed when the incoming gluon is left-handed,
at least at very large $p_T^W$, and when the $W$ spin is analyzed along
the $W$ flight direction, as measured in the partonic center-of-mass frame.
We will see in a moment that this statement is true even at lower $p_T^W$.
The purely left-handed $W$ polarization is indicated
by a long downward-pointing vertical arrow next to the $W$
in \fig{ugWdFigure}(a).

In contrast, when the incoming gluon is right-handed, as in
\fig{ugWdFigure}(b), we use \eqn{slashpolglueLout} for the gluon
polarization bi-spinor.  Now $|q^-\rangle$ enters a more complicated
spinor string, and we cannot make the $t$-channel graph vanish by a simple
choice of $q$.  (We could make the $s$-channel graph vanish if we wanted
to, but that would not help in an angular-momentum-based argument.)
In accordance with this obstruction, the outgoing $W$ now can have any of the
three possible helicities.  We will see shortly that when the $W$
transverse momentum becomes large, its polarization is dominantly
right-handed.  The indefinite $W$ polarization, but transitioning
to right-handed, is indicated by a short arrow angling upward, next to the $W$
in \fig{ugWdFigure}(b).

The reason the right-handed $W$ polarization in case (b) does not wash
out the left-handed polarization in case (a) at large transverse momenta
is because the magnitude of its squared matrix element is only
about $1/4$ the size of the one in case (a).  This smaller weighting leads to
an estimated asymptotic polarization, at very large $W$ transverse momentum,
of roughly 80\% left-handed and 20\% right-handed.
In this limit, the left-handed $W$ bosons all come from left-handed gluons,
and the right-handed ones are all from right-handed gluons.
We will see later in the paper that the actual $W$ polarizations
predicted, at transverse momenta accessible at the LHC, are remarkably
close to this asymptotic value.

\begin{figure*}[tbh]
\begin{minipage}[b]{1.\linewidth}
\includegraphics[clip,scale=.7]{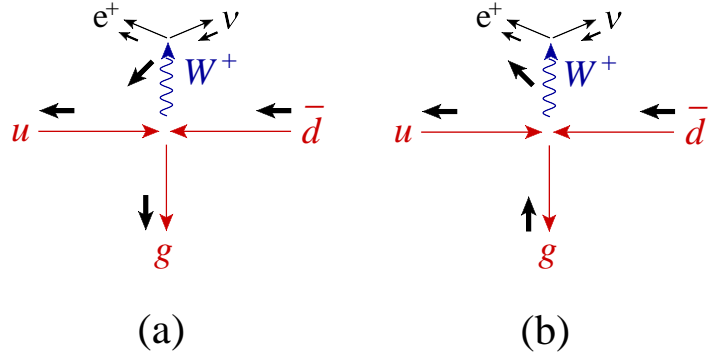}
\end{minipage}
\caption{Helicity configurations for the subprocess $u\bar{d}\to Wg$ for
(a) a right-handed outgoing gluon and (b) a left-handed one.
The directions of the $W$ spin arrows are discussed in the text.}
\label{udbarWgFigure}
\end{figure*}

\begin{figure*}[tbh]
\begin{minipage}[b]{1.\linewidth}
\includegraphics[clip,scale=.7]{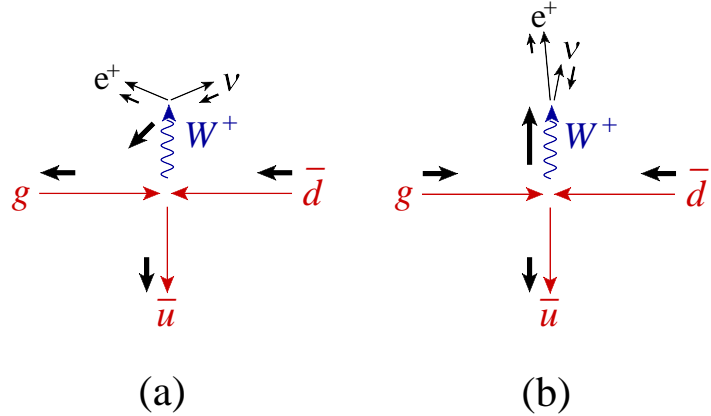}
\end{minipage}
\caption{Helicity configurations for the subprocess $g\bar{d}\to W\bar{u}$
for (a) a left-handed incoming gluon and (b) a right-handed one.
Case (b) yields a purely right-handed $W$ boson, while case (a) tends
toward left-handed at large $p_T^W$, but with a smaller weight, as
indicated by the arrow.}
\label{gdbarWubarFigure}
\end{figure*}

The leading-order amplitudes for the three subprocesses in
\eqn{subprocesshierarchy} are all related by crossing symmetry.
After decaying the $W$ to a lepton pair, $W^+ \to l^+ \nu$,
they can all be written simply in terms of spinor products,
\bea
{\cal A}^{\tree}_{\rm (a)} &\propto&
\frac{ {\spa{d}.{\nu}}^2 }{\spa{u}.{g} \spa{g}.{d}}
\quad\qquad\Rightarrow\quad\qquad
d\sigma^{\rm LO}_{\rm (a)} \,\propto\, (k_d\cdot k_\nu)^2 \,,
\label{Casea}\\
{\cal A}^{\tree}_{\rm (b)} &\propto&
\frac{ {\spb{u}.{e}}^2 }{\spb{u}.{g} \spb{g}.{d}}
\quad\qquad\Rightarrow\quad\qquad
d\sigma^{\rm LO}_{\rm (b)} \,\propto\, (k_u\cdot k_e)^2 \,,
\label{Caseb}
\eea
where we dropped coupling and propagator factors common to the two cases.
The helicity configurations (a), given by \eqn{Casea}, are depicted
in figs.~\ref{ugWdFigure}(a), \ref{udbarWgFigure}(a) and
\ref{gdbarWubarFigure}(a).  The configurations (b), given by \eqn{Caseb},
are shown in figs.~\ref{ugWdFigure}(b), \ref{udbarWgFigure}(b) and
\ref{gdbarWubarFigure}(b).

For \fig{ugWdFigure}(a), the factor of $(k_d\cdot k_\nu)^2$ in the $W$
rest frame is proportional to $(1-\cos\thstilde)^2$, where $\thstilde$
is the angle between the charged lepton and the $W$ flight direction,
as measured in the partonic center-of-mass frame. This is also the
angle between the $d$ quark and neutrino directions.  This angular
dependence implies the purely left-handed $W$ boson polarization
mentioned above.  In contrast, for \fig{ugWdFigure}(b), the factor of
$(k_u\cdot k_e)^2$ does not lead to a net left-handed polarization.
It correlates the positron direction with the incoming beam direction
rather than with the outgoing $W$ flight direction.  In the limit of
large transverse momentum, when one boosts from the parton
center-of-mass frame to the $W$ rest frame, the incoming $u$ quark and
gluon are almost parallel, and their spatial momentum adds up to give
the outgoing $d$ quark momentum.  If the scattering angle is
$90^\circ$ in the center-of-mass frame, then the magnitude of the $u$
quark momentum is precisely half that of the $d$ quark, in the $W$
rest frame.  Then the numerator factor $(k_u\cdot k_e)^2$ in
\eqn{Caseb} yields the opposite polarization from $(k_d\cdot k_\nu)^2$
in \eqn{Casea}, but only at $1/4$ the rate, {\it i.e.}~it is
proportional to $\frac{1}{4}(1+\cos\thstilde)^2$.

A partonic scattering angle of $90^\circ$ kinematically maximizes $p_T^W$
at fixed parton center-of-mass energy, {\it i.e.}~at fixed $x_1x_2$
for $u(x_1)g(x_2)\to W^+ d$.  However, the matrix element prefers
a smaller scattering angle (the $d$ quark more parallel to the incoming
gluon), while the parton densities prefer $x_1>x_2$, which further skews
the preferred kinematics.  Hence the 80\% value for $f_L$ that is implied
by $90^\circ$ scattering is just an estimate for asymptotically
large $p_T^W$.

At finite $W$ transverse momentum, we find that the polarization fractions
for $90^\circ$ $ug\to W d$ scattering in case (b) are,
\begin{equation}
f_L = \frac{1}{4} (1-\cos\theta_u)^2 \,, \hskip 1 cm 
f_R = \frac{1}{4} (1+\cos\theta_u)^2 \,, \hskip 1 cm 
f_0 = \frac{1}{2} \sin^2\theta_u \,.
\label{fb}
\end{equation}
where $\theta_u$ is the angle that the $u$ quark makes with the $d$ quark
in the $W$ rest frame.  In terms of the boost of the $W$ boson in
the partonic center-of-mass frame, $\gamma = E_W/M_W$, it satisfies,
\be
\sin\theta_u = \frac{1}{\gamma} \,.
\label{sinthetau}
\ee
At $90^\circ$, the overall weighting of this helicity configuration, with
respect to case (a), is $1/(4\cos^2\theta_u)$.  From these relations
one can estimate the LO polarization fractions from this subprocess
at finite $p_T^W$.

The subprocess $u\bar{d} \to W^+g$ shown in \fig{udbarWgFigure} is subdominant
to $ug\to W^+d$, but it can be analyzed similarly using \eqns{Casea}{Caseb}.
In both cases the decay leptons are correlated with the beam direction.
At large $W$ transverse momenta, case (a) yields mainly
left-handed $W$ bosons, while (b) yields mainly right-handed ones
(as indicated by the arrows next
to the $W$s in the figure).  For $90^\circ$ scattering, the
two cases for $u\bar{d} \to W^+g$ cancel,
and there is no net left-handed polarization from this subprocess.
Finally, the subprocess $g\bar{d} \to W^+\bar{u}$ is shown in
\fig{udbarWgFigure}.  It produces a net right-handed $W$ polarization,
from configuration (b), because the factor $(k_u\cdot k_e)^2$ is proportional
to $(1+\cos\thstilde)^2$.  However, it is suppressed compared to
the dominant source of left-handed polarization in \fig{ugWdFigure}(a),
because $u(x) \gg \bar{d}(x)$ except at quite small $x$.

We have argued that the $W$ polarization should reach about 80\% left-handed
and 20\% right-handed at asymptotically large $W$ transverse momentum.
However, there are a number of reasons why the left-handed fraction
should be smaller at finite $p_T^W$:
\begin{enumerate}
\item The mainly right-handed configuration in \fig{ugWdFigure}(b)
competes better against the pure left-handed one in \fig{ugWdFigure}(a)
for smaller $p_T^W$.
\item The 100\% left-handed fraction found in \fig{ugWdFigure}(a) was
analyzed with respect to a $W$ flight direction measured from the partonic
center-of-mass frame, but the conventional definition is from the $pp$
center-of-mass frame, which differs whenever the $u$ quark and gluon
momentum fractions are not identical.
\item The subdominant $u\bar{d}$ and $g\bar{d}$ channels dilute the
polarization.
The dilution decreases as $x$ increases, {\it i.e.} as $p_T^W$
increases (or as other measures of the hardness of the event increase,
such as the scalar transverse energy $H_T$ for a $W+$\,multi-jet event).
\item While QCD corrections are generally expected to be small ---
and we will confirm this expectation in this paper --- in principle
they can affect the $W$ polarization fractions.
\end{enumerate}
Note that the third remark suggests that the left-handed fraction
for $W^+$ bosons should be a bit larger than the left-handed fraction
for $W^-$ bosons, at a given $p_T^W$.  This property should hold
because in the proton $u(x) > d(x)$, which allows the dominance
of $ug\to W^+d$ over $g\bar{d}\to W^+\bar{u}$ to set in before that
of $dg \to W^-u$ over $g\bar{u}\to W^-\bar{d}$.

In refs.~\cite{W3jDistributions,Polarization}, left-handed $W$
polarization effects were shown to be large in \Wjj-jet and
\Wjjj-jet production, for moderate to large $p_T^W$.
The generic kinematics for these processes are
quite complicated, and we don't fully understand why the polarization
is so large here.  For very large $W$ transverse momentum, the
configuration preferred by the fast-falling parton distributions is
one in which the $W$ recoils against a cluster of jets with relatively
small invariant mass.  In this limit, the multi-parton amplitudes can
be factorized~\cite{W3jDistributions} into the ones for a $W$
recoiling against a single parton, multiplied by collinear or
multi-collinear QCD splitting amplitudes.  Because the QCD splitting
amplitudes are invariant under parity, one can use the same argument
given above for the case of \Wpj-jet production (or alternatively, for
$W$ production at a finite transverse momentum, with no explicit jet
requirement). 
However, for moderate $W$ transverse momenta, close to the jet $p_T$
threshold, the configuration in which a $W$ boson recoils against a
small invariant-mass cluster of jets should be fairly rare, and so this
argument would not apply.

\begin{figure*}[tbh]
\begin{minipage}[b]{1.\linewidth}
\includegraphics[clip,scale=.5,angle=270]{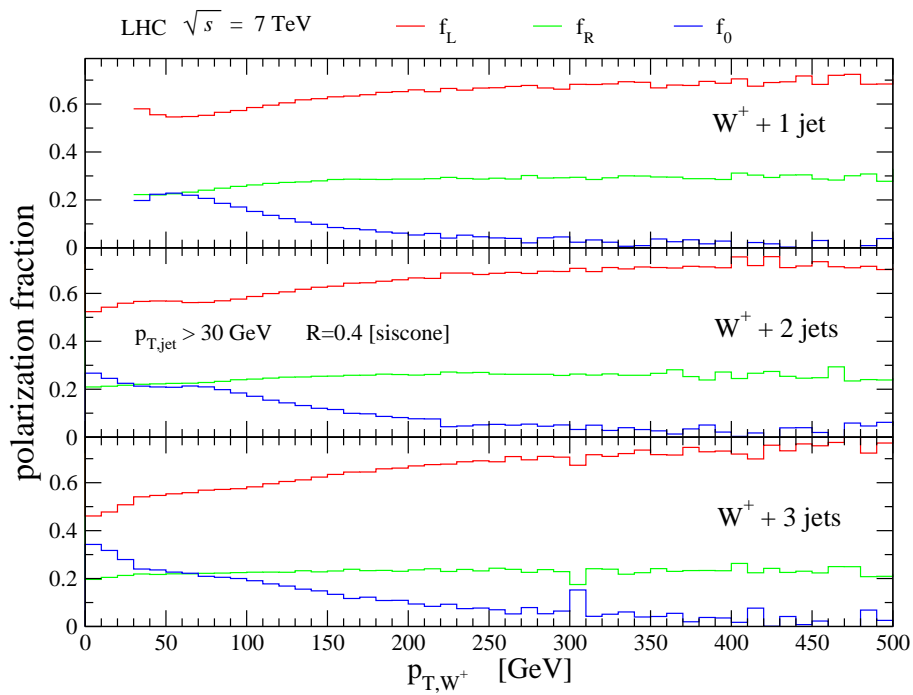} 
\end{minipage}
\caption{A comparison of the polarization fractions for \Wpjjja-jet
  production at LO, illustrating the insensitivity of the
  polarization fractions to the number of jets. 
   The top panel shows
  the polarization fractions for \Wpj-jet production, the middle panel
  \Wpjj-jet production, and the third panel \Wpjjj-jet production. In
  each panel (except at low $p_T^W$) the top curve (red) gives $f_L$,
  the middle curve (green), $f_R$, and the bottom one (blue), $f_0$. }
\label{WpComparisonFigure}
\end{figure*}

\begin{figure*}[tbh]
\begin{minipage}[b]{1.\linewidth}
\includegraphics[clip,scale=.6,angle=270]{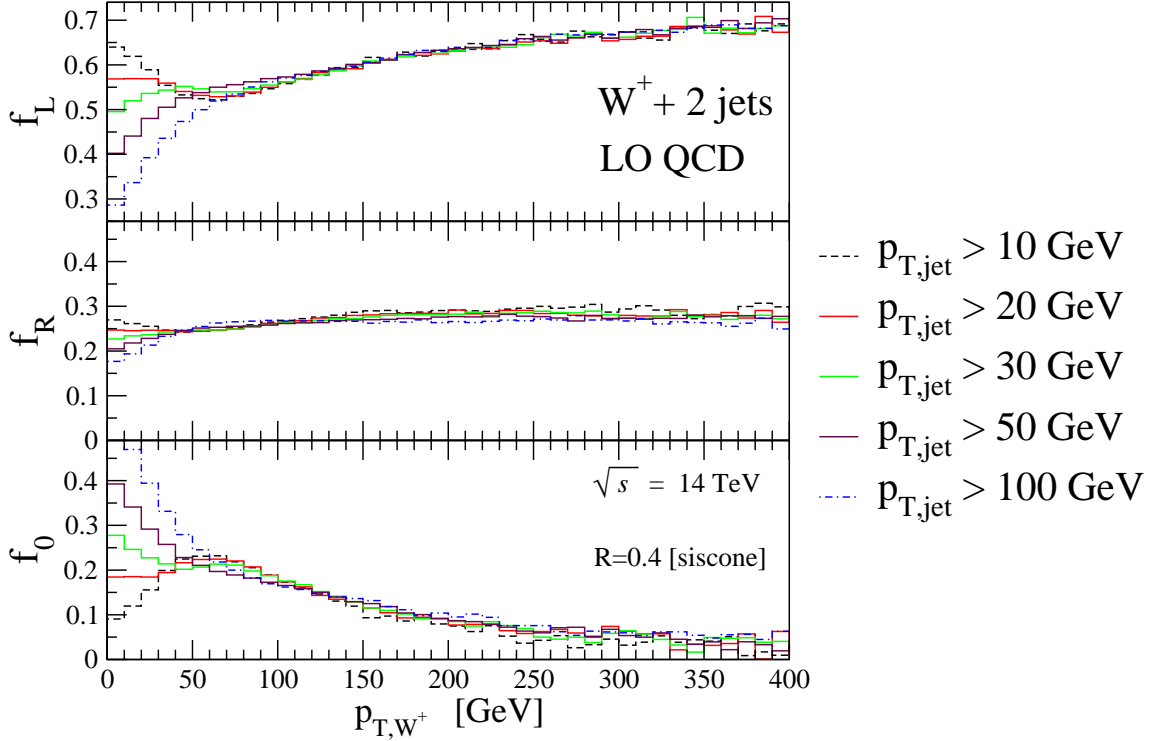}
\end{minipage}
\caption{The polarization fractions in \Wjj-jet production as a function of
$p_T^W$, with jet cuts of $p_{T,\rm jet} > 10,20,30,50,100$ GeV, at
LO. The dashed (black) curve corresponds to a 10 GeV cut and the
dot-dashed (blue) curve corresponds to a 100 GeV cut; the others are in
between.  At small $p_T^W$ the 10 GeV cut gives the largest $f_L$ and $f_R$
fractions, while the 100 GeV cut gives the smallest $f_L$ and $f_R$
fractions.  Above $p_T^W = 50$ GeV
there is little sensitivity to the jet cuts. 
The center-of-mass energy is $\sqrt{s} = 14$~TeV.  }
\label{JetCutDependence}
\end{figure*}

\begin{figure*}[tbh]
\begin{minipage}[b]{1.\linewidth}
\includegraphics[clip,scale=.5,angle=270]{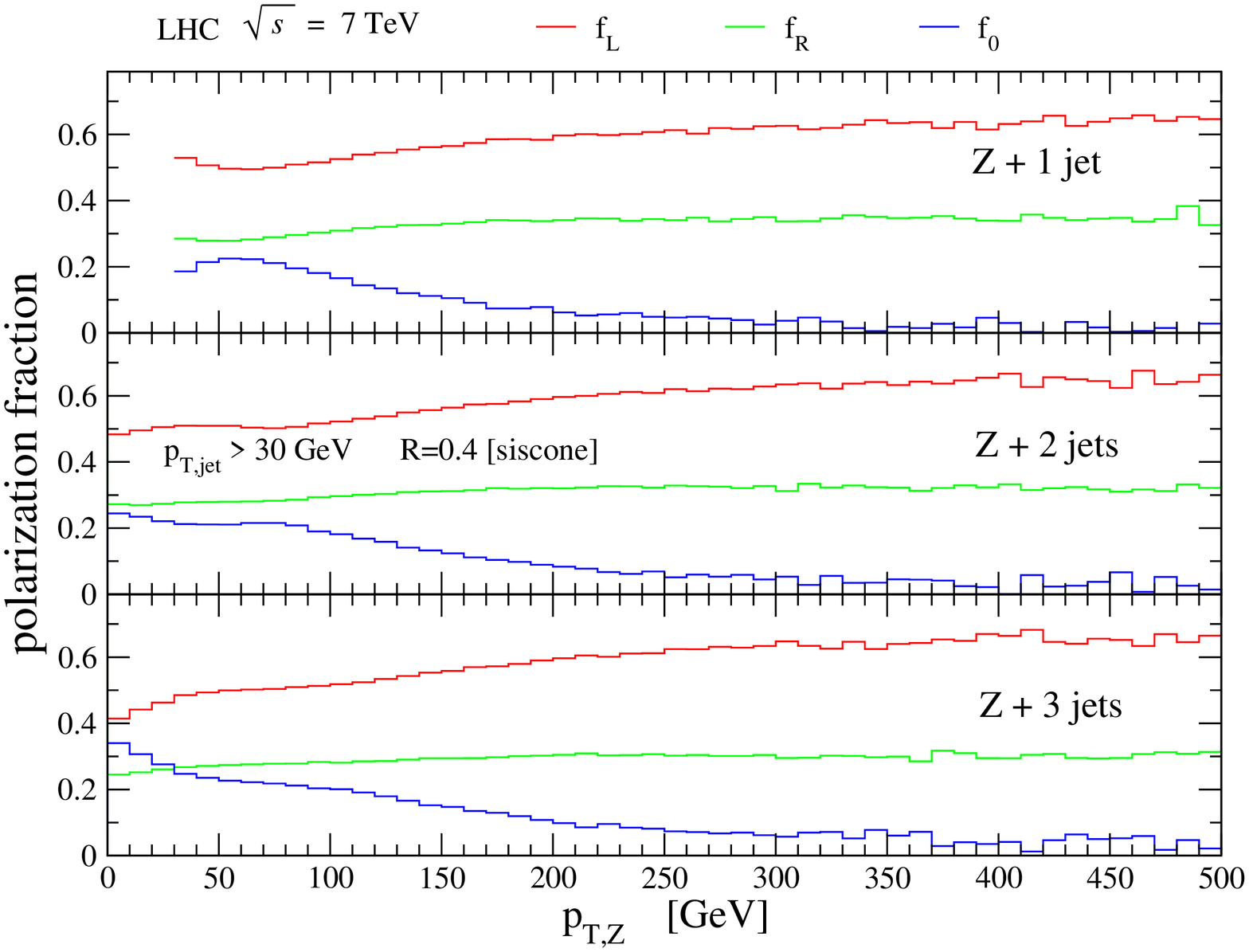} 
\end{minipage}
\caption{The same comparison as in \fig{WpComparisonFigure}, except
  for $Z$ bosons.}
\label{ZComparisonFigure}
\end{figure*}

In any case, it is not difficult to confirm using standard Monte Carlo
programs that the average degree of polarization at large $p_T^W$ is
rather insensitive to the number of jets.  In \fig{WpComparisonFigure} we
compare the polarization fractions as a function of $p_T^W$ for
\Wpjjja-jet production at (unshowered, fixed-order) LO, using the
\SHERPA{} package.  For these plots we imposed a $p_T > 30$~GeV cut on
the jets, using the \SISCone{} jet algorithm~\cite{SISCONE} with
$R=0.4$ and the CTEQ6L1~\cite{CTEQ6M} parton distribution set.  Beyond
low vector-boson $p_T$, the three cases are remarkably similar, with $f_L$
reaching 70\% at high vector-boson $p_T$.  
(The sharp cutoff of events below $30$
GeV for the \Wj-jet case is an artifact of LO QCD which constrains the
$W$ to balance the $p_T$ of the jet, required to exceed $30$ GeV.)  The
insensitivity of the polarization fractions to the number of jets
holds just as well at NLO.  Another interesting feature is the
insensitivity of the $W$ polarization to the jet cuts for
$p_T^W>50$~GeV.  This feature is illustrated in \fig{JetCutDependence}
for \Wpjj-jet production, varying jet cuts from 10 to 100 GeV.  The
setup in this illustration is the same as for \fig{WpComparisonFigure}, except
the $pp$ center-of-mass energy is $\sqrt{s} = 14$~TeV, instead of
$\sqrt{s} = 7$ TeV.  The insensitivity to the number of jets,
and to the jet cuts, also holds for $W^-$ production.

Interestingly, $Z$ bosons behave similarly at the LHC, achieving a
slightly lower polarization, but with $f_L$ still reaching above 60\%,
as shown in \fig{ZComparisonFigure}.
This lowering happens because $Z$ bosons do couple to
right-handed quarks, and right-handed initial-state
quarks lead to reversed vector-boson polarization.
(The $u$ quarks producing $Z$ bosons
are 16\% right-handed, 84\% left-handed, while the $d$ quarks are
only 3\% right-handed, for $\sin^2\theta_W = 0.23$.)
However, the $Z$ polarization is more difficult to 
measure because it is less efficiently analyzed by
$Z \rightarrow l^+ l^-$, as the $Z$ coupling to the leptons
is close to equally left- and right-handed.  The analyzing power is
only about 15\% in the leptonic $Z$ decay, versus 100\% for $W\to l\nu$.

The analysis of high-$p_T$ $W$ production at
the Tevatron differs, because it
is a $p\bar{p}$ collider.
Here the weighting of the partonic subprocesses represented in 
figs.~\ref{ugWdFigure}, \ref{udbarWgFigure} and \ref{gdbarWubarFigure}
is quite different, with \fig{udbarWgFigure} ($u\bar{d}\to W^+g$)
much more important, because both incoming quarks are now valence quarks.
Because this subprocess does not lead to a large
net left-handed $W$ polarization, 
and because the $ug\to W^+d$ and $g\bar{d}\to W^+\bar{u}$ subprocesses
are quite close in magnitude (and lead to opposite left {\it vs.}~right
polarization), we expect
$f_L \approx f_R$ for both $W^+$ and $W^-$.   The CP invariance
of the initial state implies that $f_L(W^\pm)=f_R(W^\mp)$, neglecting
CP violation, and as long as
the acceptances are symmetric with respect to reversing the $p$ and $\bar{p}$
directions.  However, one can increase the polarization of
$W^+$ bosons with nonvanishing $p_T$ at the Tevatron, by requiring them
to be in the forward hemisphere (the proton direction).
Such a cut will increase the contribution
of $ug\to W^+d$ relative to $g\bar{d}\to W^+\bar{u}$.


\section{Defining and computing polarization}
\label{MethodsSection}

\subsection{Polar angle dependence}

The angular distribution of the leptonic $W$ decay products in the $W$
rest frame is given by a standard helicity analysis, predicated on the
spin-1 nature of the $W$, and the fact that the fermions
(anti-fermions) to which it decays are purely left-handed
(right-handed).  First we consider the distribution in the polar angle
$\ths$, after integrating over the azimuthal angle $\phis$.  We boost
from the lab frame to the $W$ rest frame.  In this frame, we define
$\ths$ to be the angle between the $W$ flight direction, as observed in
the lab frame, and the charged lepton.
This angle takes values in the interval $[0,\pi]$.  The distribution in
$\ths$, in terms of the polarization fractions $f_L, f_R$ and $f_0$,
is~\cite{ESW},
\begin{equation}
{1\over\sigma} {d\sigma\over d\cos\ths}
\ =\  \frac{3}{8} (1 \mp \cos\ths)^2 \,f_L
    + \frac{3}{8} (1 \pm \cos\ths)^2 \,f_R
    + \frac{3}{4} \sin^2\ths \, f_0 \,,
\label{dcdist}
\end{equation}
where the upper sign is for $W^+$ and the lower sign for $W^-$.
The normalizations are chosen so that
\begin{eqnarray}
	\int_{-1}^{1} d\cos\ths {1\over \sigma} \, 
        \frac{d\sigma}{d\cos\ths} = f_L + f_0 + f_R = 1 \,.
\end{eqnarray}
For a left-handed $W^+$ the decay amplitude must vanish at $\ths = 0$
because angular momentum would be violated in a decay to a
forward-going right-handed anti-lepton and a backward-going
left-handed lepton.  This explains why the $f_L$ term is proportional
to $(1-\cos\ths)^2$ for $W^+$.  Similarly, for a right-handed $W^+$,
the decay must vanish for $\ths = \pi$, explaining the factor of
$(1+\cos\ths)^2$ multiplying $f_R$.  Decays of the longitudinal mode
are forbidden at both $\ths = 0$ and $\ths = \pi$, explaining the
$\sin^2\ths$ behavior.  The $W^-$ case behaves oppositely because the
charged lepton is now left-handed rather than right-handed.

In \eqn{dcdist}, $\sigma$ can be a differential cross section. For
example, \eqn{dcdist} is just as valid if we replace 
\begin{equation}
\sigma \rightarrow \frac{d\sigma}{d p_T^W} \,.
\end{equation}
In fact, any differential cross section that does not depend on the
kinematics of individual leptons can be used.  Inserting such a
distribution into \eqn{dcdist} allows us to define
polarization fractions $f_L$, $f_R$ and $f_0$ as a function of the
$W$ boson kinematics, number of jets, and so forth.

We can define the expectation of an observable $g(\ths)$ via
\begin{eqnarray}
\langle g (\ths) \rangle \equiv  
\int_{-1}^{1} g(\ths) \,
 \frac{1}{\sigma} \frac{d\sigma}{d\cos \ths} \, d\cos\ths  \,.
\end{eqnarray}
In particular, the expectation value $\langle \cos\ths \rangle$ is
\begin{eqnarray}
\langle \cos\ths\rangle=
 \int_{-1}^{1} \cos\ths 
       \frac{1}{\sigma} \frac{d\sigma} {d\cos\ths} \, 
  d\cos \ths 
 = \mp \frac{1}{2} (f_L - f_R) 
 = \pm \biggl( \frac{1}{2} - f_L -  \frac{1}{2} f_0 \biggr)\,.
\label{CosTheta}
\end{eqnarray}
We can obtain other moments similarly, such as
\begin{eqnarray}
&&\langle \cos^2\ths\rangle = \frac{2}{5} - \frac{1}{5} f_0 \,,\\
&&\langle \cos^4\ths\rangle = \frac{9}{35} - \frac{6}{35} f_0\,.
\end{eqnarray}
Solving for the longitudinal fraction $f_0$ gives,
\begin{equation}
f_0 = 2 - 5 \langle \cos^2\ths\rangle\,,
\label{F0c2}
\end{equation}
or
\begin{equation}
f_0 = \frac{3}{2} - \frac{35}{6} \langle \cos^4\ths \rangle\,.
\label{F0c4}
\end{equation}
We compute lepton decay distributions numerically by Monte Carlo sampling,
and we accumulate several different moments at once.
Using the two formul\ae{}~(\ref{F0c2}) and (\ref{F0c4}) should give the same
answer for $f_0$.

Finally, by plugging in \eqn{F0c2} into \eqn{CosTheta} we can solve for
the left- and right-handed polarization fractions,
\begin{eqnarray}
f_L = - \frac{1}{2} \mp \langle \cos\ths\rangle 
      + \frac{5}{2}   \langle \cos^2\ths\rangle \,, \nn \\
f_R = - \frac{1}{2} \pm \langle \cos\ths\rangle 
      + \frac{5}{2}   \langle \cos^2\ths \rangle \,,
\end{eqnarray}
where again the top sign is for $W^+$ and the bottom sign is for $W^-$.


\subsection{Inclusion of the azimuthal angle}

As mentioned in \sect{IntroductionSection}, detector effects such as
finite resolution, acceptance and reconstruction efficiency distort
angular distributions, so that the extracted polarization fractions are
sensitive to how the cross section
depends on the azimuthal angle $\phis$ as well as $\ths$.  We therefore give
the complete dependence of the cross section on $\ths$ and $\phis$.
Similar angular decompositions may be found in
refs.~\cite{CollinsSoper,LamTung,Hagiwara1,Mirkes,MirkesOhnemus,Hagiwara2}.
One important difference is that we do not use the Collins-Soper
frame~\cite{CollinsSoper} but rather define angles using the $W$ boson
flight direction.  As we saw in the previous section, this definition
reveals the left-handed nature of the produced $W$ bosons quite
cleanly.

\begin{figure*}[tbh]
\begin{minipage}[b]{1.\linewidth}
\includegraphics[clip,scale=.5]{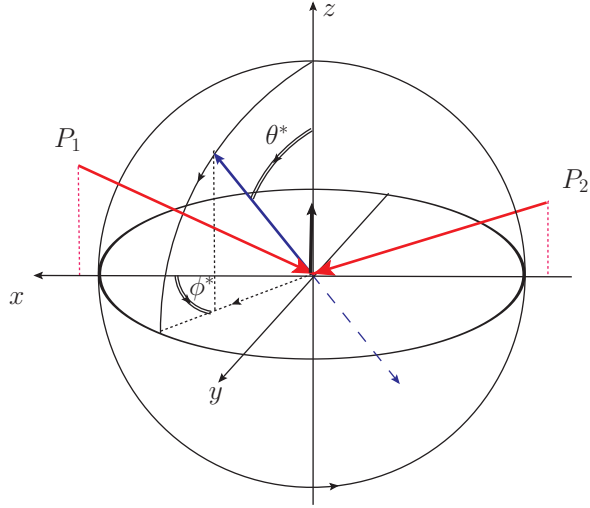}
\end{minipage}
\caption{ The lepton decay angles in the $W$ rest frame. 
  The original $W$ flight direction,
defining the $z$-axis, is represented by a dark (black) arrow. 
The protons, represented by angled (red) arrows pointing at the
origin, lie in the $(x,z)$-plane.  The
momenta of the charged lepton and the neutrino are given by the solid
and dashed blue arrows, respectively.
The decay angle $\ths$ is measured with respect to the $z$ axis. 
The origin of the azimuthal angle $\phis$ is along the $x$ axis, which
lies in the plane defined by the proton momenta. The positive $x$ axis
points in the direction of motion of 
the proton with the smaller angular separation
from the $z$ axis ($P_2$ here).  The coordinate system is right-handed,
which defines the direction of the $y$ axis.
}
\label{DecayAngles}
\end{figure*}

As before, we boost from the lab frame to the $W$ rest frame. 
The $W$ flight direction defines the $z$-axis.  The $(x,y)$-plane is
orthogonal to the $z$-axis, and $(x,y,z)$ form a right-handed
coordinate system (see \fig{DecayAngles}).
The azimuthal angle $\phis$ takes values in $[0,2\pi)$ and is equal to 0
in the positive $x$ direction, $\pi/2$ in the positive $y$ direction.
The $x$-axis is defined by the intersection of the plane spanned by
the two proton momenta with the $(x,y)$-plane.  Finally, the
orientation of the positive $x$-axis is defined using the proton momenta.
The positive $x$-axis is defined~\cite{CMSMeasurement} to point in the
direction of the proton with the smaller angular separation from the
$z$-axis ($P_2$ in the case shown in \fig{DecayAngles}).

We consider the decay distribution of the $W$ boson at finite $p_T^W$,
in terms of the lepton angles defined in \fig{DecayAngles}. Following
refs.~\cite{CollinsSoper,LamTung,Hagiwara1,Mirkes,MirkesOhnemus,Hagiwara2},
we decompose the cross section as
\begin{eqnarray}
\frac{1}{\sigma} \frac{d\sigma}{d(\cos{\ths})d\phis}&=&
 \frac{3}{16\pi} \Bigl[
 (1+\cos^2{\ths})+
 A_0\,\frac{1}{2}(1-3 \cos^2{\ths}) +
 A_1\,\sin{2\ths}\cos{\phis} \nn \\&& \null 
+A_2\, \frac{1}{2}\sin^2{\ths}\cos{2\phis}+
 A_3\, \sin{\ths}\cos{\phis}+
 A_4\, \cos{\ths}
 \nn \\&& \null
+A_5\, \sin{\ths}\sin{\phis} +
 A_6\, \sin{2\ths}\sin{\phis}+
 A_7\, \sin^2{\ths}\sin{2\phis}
\Bigr]\,.
\label{AiSDecomposition}
\end{eqnarray}
Here we define the expectation value as 
\begin{eqnarray}
\langle f(\ths,\phis)\rangle
=\int_{-1}^1  d(\cos{\ths}) \int_0^{2\pi} d\phis \,
\frac{1}{\sigma} \frac{d\sigma}{d(\cos{\ths})d\phis} \,
 f(\ths,\phis)\,.
\end{eqnarray}
As before, $\sigma$ can be any differential cross section that does
not depend on the individual lepton kinematics.

\begin{figure*}[tbh]
\begin{minipage}[b]{1.\linewidth}
\includegraphics[clip,scale=.4]{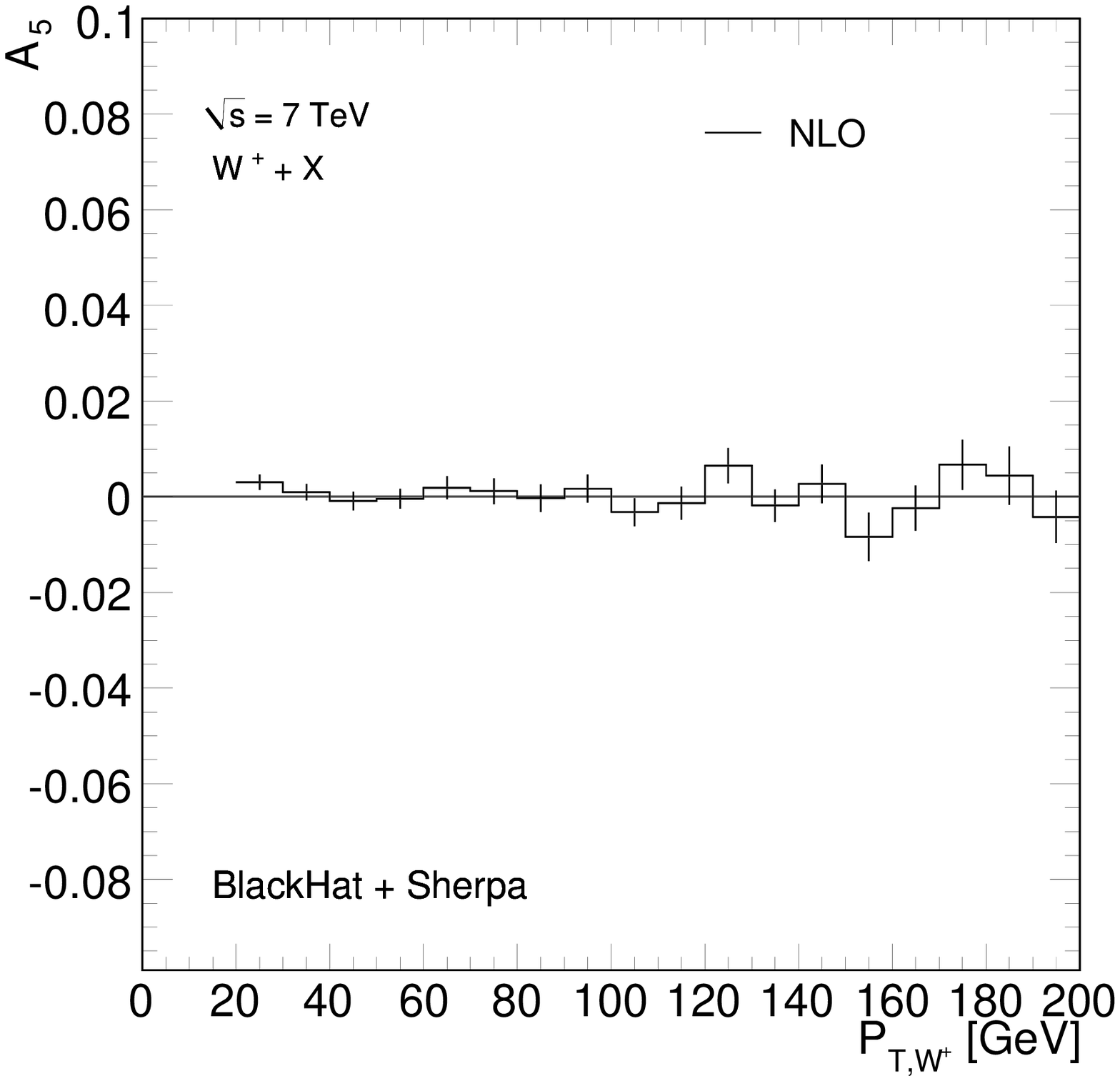}
\includegraphics[clip,scale=.4]{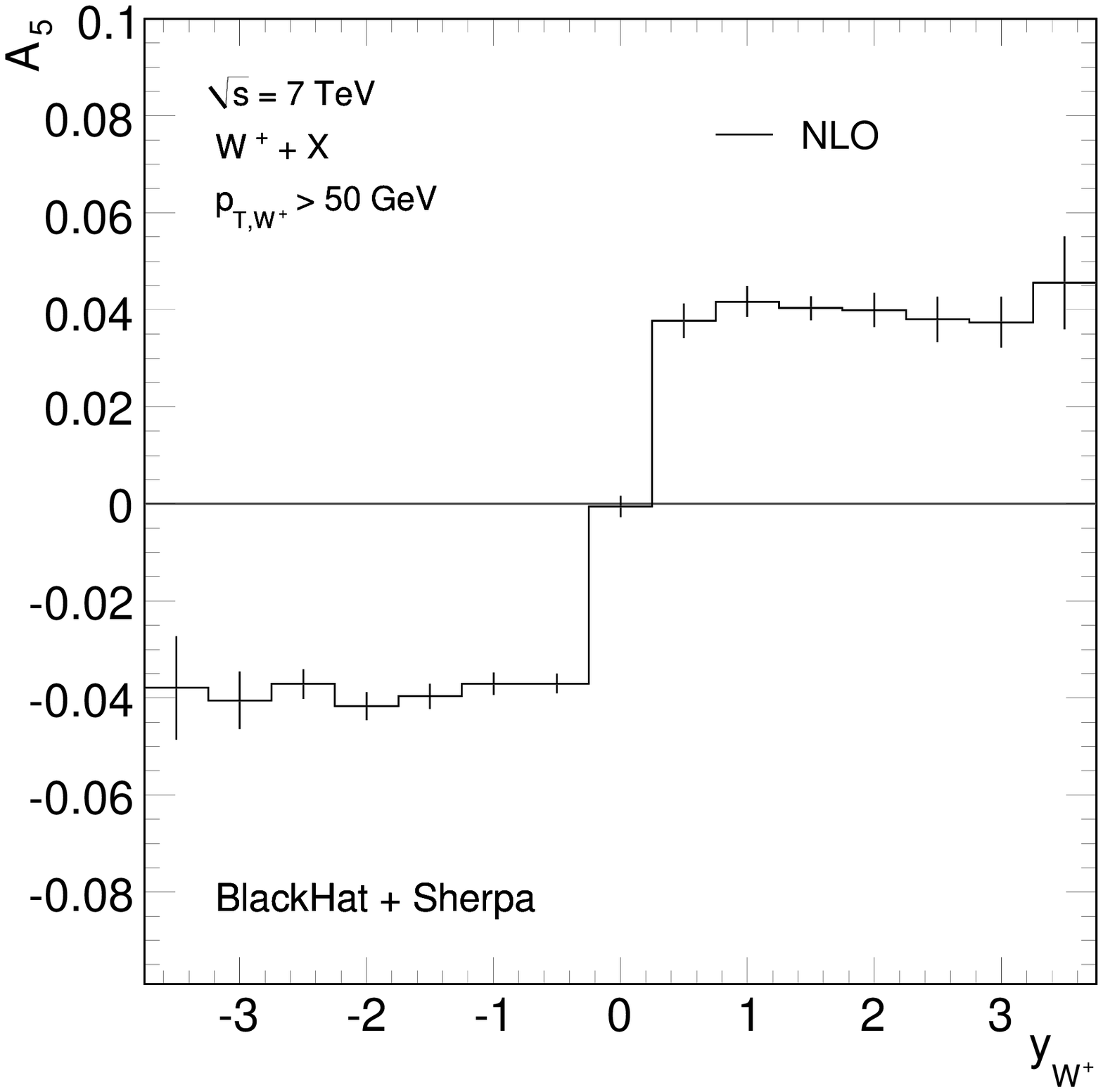}
\end{minipage}
\caption{The left panel shows $A_5$ as a function of $p_T^W$ for 
$W^+$ with any number of jets at the LHC at NLO;
the result is very small, if not vanishing.  The
right panel shows $A_5$ as a function of the rapidity, 
with a cut of  $p_T^W > 50$ GeV imposed.
The thin vertical lines give the integration errors.
}
\label{A5Figure}
\end{figure*}

Following similar logic as for the previous case, in which the 
azimuthal angle has been integrated out, we may extract the
angular coefficients $A_i$ directly in terms of expectation values,
\begin{eqnarray}
A_0&=& 4 - 10 \, \langle \cos^2{\ths} \rangle\,, 
\hskip 1.2 cm 
A_1=\langle 5 \sin{2 \ths} \cos{\phis} \rangle\,,  \hskip 1.2cm 
A_2=\langle 10\sin^2{\ths} \cos{2 \phis}\rangle \,, \nn \\
A_3&=&\langle 4\sin{\ths} \cos{\phis}\rangle\,, \hskip 1.25 cm 
A_4=\langle 4\cos{\ths}\rangle\,, \hskip 2.45 cm 
A_5=\langle 4\sin{\ths} \sin{\phis}\rangle\,, \nn \\
A_6&=&\langle 5\sin{2 \ths} \sin{\phis}\rangle\,, \hskip 1.1 cm 
A_7=\langle 5\sin^2{\ths} \sin{2 \phis}\rangle \,.
\end{eqnarray}

For either LO or ME+PS, the coefficients $A_5, A_6, A_7$
vanish, because the functions they multiply
are odd under a ``naive'' time-reversal symmetry,
as well as under parity, either of which maps $\phis\to-\phis$.
They do receive contributions at NLO in QCD from the absorptive part
of one-loop amplitudes~\cite{Hagiwara1,Hagiwara2}.  However, the $p_T^W$
distribution of these contributions is highly suppressed at the LHC.
The left panel of \fig{A5Figure} shows that the $A_5$ coefficient is
well below 0.01, and indeed is consistent with zero, within
integration errors due to sampling fluctuations.  However, from the
right panel of \fig{A5Figure}, we see that this suppression is simply
due to a cancellation between the forward and backward
regions in rapidity.  The rapidity distributions for $A_6$ and $A_7$
are smaller than the one for $A_5$.
An analogous computation for the parity-odd coefficients,
both in $p\bar{p}$ and $pp$ collisions up to 20 TeV, was given in
ref.~\cite{Hagiwara1}, but using the Collins-Soper frame.
Similar results hold
when the $W^+$ is replaced by a $W^-$.  In the remaining
part of this paper, we will not distinguish between the forward and
backward rapidity regions, and we will not discuss
$A_5$, $A_6$ and $A_7$ further.

If we integrate the combined distribution in $\ths$ and $\phis$,
\eqn{AiSDecomposition}, over the azimuthal angle $\phis$,
we should recover \eqn{dcdist} for the polar-angle distribution.
Carrying out the integration, we obtain
\begin{eqnarray}
\frac{1}{\sigma}
 \frac{d\sigma}{d(\cos{\ths})}&=&
\frac{3}{8} \left[
 (1+\cos^2{\ths})+A_0\,\frac{1}{2}(1-3 \cos^2{\ths})+
A_4\, \cos{\ths}\right]\,.
\end{eqnarray}
Comparing to \eqn{dcdist}, we see that the polarization fractions are
given in terms of the $A_i$ as
\begin{equation}
f_L = \frac{1}{4} (2 - A_0 \mp A_4)\,, \hskip 1 cm 
f_R = \frac{1}{4} (2 - A_0 \pm A_4)\,, \hskip 1 cm 
f_0 = \frac{1}{2} A_0 \,,
\end{equation}
with the top sign for $W^+$, the bottom sign for $W^-$.
We also have $f_L-f_R = \mp A_4/2$.


\section{Results}
\label{ResultsSection}

In this section we present our results for the polarization coefficients.
Prior to presenting them, we summarize our calculational setup based
on \BlackHat{}~\cite{BlackHatI} and \SHERPA~\cite{Sherpa}.  We will
present results for both NLO parton-level QCD and the ME+PS
framework.  For reference, to gauge the size of
higher-order QCD corrections, we also present results at LO
fixed-order parton-level. We find that the polarization fractions
are quite insensitive to variations of a common renormalization and
factorization scale, so varying the scale does not offer a reasonable
estimate of the theoretical uncertainty.  Instead, we use the difference
between the NLO and ME+PS predictions as a measure of the uncertainty.
The two frameworks have competing strengths.  The NLO result benefits
from true virtual corrections and a better cancellation of scale
dependence in the overall cross section.  The ME+PS result includes the
effects of radiating multiple soft gluons.  The similarity of
both predictions gives us confidence that further refinement
in the order (fixed or logarithmic) perturbative expansion
will not lead to large changes in these observables.


\subsection{Setup}

We use \SHERPA\ for several different purposes: computing the basic
fixed-order LO predictions; generating the ME+PS predictions; and for
providing the phase-space integration framework as well as the
real-emission contributions to the fixed-order NLO predictions.  ME+PS
event samples are produced according to the technique described in
ref.~\cite{MEPS}. This method combines two essentially different
approaches to perturbative QCD, hard matrix element calculations,
which are exact at some fixed perturbative order (LO in our case) and
parton showers, which resum logarithmic corrections due to
Bremsstrahlung effects.  The parton shower employed to this end in
\SHERPA{}~\cite{CSShower} is based on Catani-Seymour dipole
factorization~\cite{CS}. In contrast to earlier parton showers, the
model inherently respects QCD soft color coherence, as the eikonal
factors associated with soft-gluon emission off a color dipole are
exactly mapped onto two dipole functions, which differ only in the
assignment of emitter and spectator partons. Additionally, the model allows
the unambiguous identification of a recoil partner for partons that
are shifted off mass-shell in the splitting process (the ``mother''
partons), thereby eliminating one of the major sources of uncertainty
in earlier schemes for parton evolution.  As the observables presented
below should be insensitive to hadronization effects, ME+PS results are
presented at the parton level.  We match to matrix elements containing up
to three final-state partons, and use 15~GeV for the merging cut.
We use the MSTW08 NLO parton distributions~\cite{MSTW}.  An identical
simulation, but using the MSTW08 LO set, gives very similar results,
except for the total cross section of the event sample.

To obtain the NLO results we use \BlackHat{} in
conjunction with \SHERPA.  We use the same basic setup employing on-shell
methods, as in earlier computations of \Wjjjx-jet and \Zgamjjj-jet
production~\cite{BHPRL,W3jDistributions,BHZ3j}. For $W$ production
with two or fewer tagged jets,
\BlackHat{} uses analytic formulas for the virtual
contributions~\cite{Z4Partons}.  For \Wjjj-jet production
(as in \fig{ChargedRatioFigure}) \BlackHat{}
evaluates these contributions numerically.
The remaining NLO ingredients, the
real-emission and dipole-subtraction terms~\cite{CS}, are computed by
\AMEGIC{}~\cite{Amegic}, part of the \SHERPA{} package~\cite{Sherpa}.
We also use \SHERPA{} to perform phase-space integration using QCD
antenna structures~\cite{AntennaIntegrator}.

In all cases, we include the full $W$ Breit-Wigner resonance and
decays to leptons retain all spin correlations. Except where noted we
use the MSTW08 NLO parton distribution functions~\cite{MSTW}.  For LO
we use the MSTW08 LO set.  Following refs.~\cite{BHZ3j,BHW4j}, we use
half the partonic total transverse energy, $\HTpartonicp/2$ as our
reference renormalization and factorization scale choice for LO and
NLO. (We define $\HTpartonicp \equiv \sum_j p_T^j + E_T^W$, where the sum runs
over all final-state partons $j$ and $E_T^W \equiv
\sqrt{M_W^2+(p_T^W)^2}$.  The $W$ transverse energy $E_T^W$ is
used, instead of the lepton sum $E_T^l+e_T^\nu$, to prevent the 
scale choice from biasing the leptonic 
angular variables.)  We found very similar NLO results using a
second, CKKW-style scale choice~\cite{Matching}.  The Standard Model
parameters are the same as in ref.~\cite{W3jDistributions}, except the
value of $\alpha_s$ is set to that employed by MSTW08.

We used \SHERPA{} version 1.3.0~\cite{SherpaCode}.
The virtual matrix elements we employed are available in 
ref.~\cite{Z4Partons}, except for the \Wjjj{}-jet matrix elements used 
in \fig{ChargedRatioFigure}, which are computed
numerically by \BlackHat{}.


\subsection{Polarization predictions}

\def\hn{\hskip -.3 cm}
\begin{table*}
\vskip .4 cm
\begin{tabular}{||c||c|c|c||c|c|c||}
\hline
 & $W^+$ NLO & $W^+$ ME+PS & $W^+$ LO &  $W^-$  NLO & $W^-$ ME+PS & $W^-$ LO  \\
\hline
$f_{\! L}$  & $\; 0.554\;$  & $\; 0.548\;$ &  $\; 0.556 \;$ 
            & $\; 0.528\;$  & $\; 0.521\;$ &  $\;0.523 \;$ \\
\hline
$f_{\! R}$  & $0.246$ & $0.265$ & $0.246$
            & $0.279$ & $0.300$ & $0.287$  \\
\hline
$f_{\! 0}$  & $0.200$ & $0.187$ & $0.198$ 
            & $0.193$ & $0.179$ & $0.190$  \\
\hline 
\end{tabular} 
\caption{The polarization fractions for $W$ production with $p_T^W>50$~GeV
and no restrictions on either the $W$ rapidity or the number of
associated jets.
\label{PolarizationFractionTable} 
}
\end{table*}

In \tab{PolarizationFractionTable} we give our predictions
for the polarization
fractions $f_L, f_R$ and $f_0$ at the LHC for both $W^+$ and $W^-$.
The $W$ bosons are required to have $p_T^W>50$~GeV,
but there is no cut on their rapidity, and no explicit jet requirements
are imposed.  We show predictions using NLO, ME+PS and LO.
The LO prediction is the least reliable of the
three, and is given only for reference purposes, to show the effect of
higher-order QCD corrections.  
This table makes clear that {\it both\/} $W^+$ and $W^-$ bosons
are predominantly left-handed.  Because the polarization fractions are
normalized by the cross section, and because we treat
the scale dependence in a correlated fashion, the renormalization- and
factorization-scale dependence is very small,
under 2\% at NLO for all fractions. It is even smaller at LO, but only
because the running of the coupling completely drops out from the
polarization fractions.  A more sensible estimate of the theoretical
uncertainty is the difference between NLO and ME+PS.
\Tab{PolarizationFractionTable} shows that this
difference is under 10\% for the polarization fractions.

\begin{table*}
\vskip .4 cm
\begin{tabular}{||c||c|c|c||c|c|c||}
\hline
 & $\;W^+$ NLO $\;$ & $\;W^+$ ME+PS $\;$ & $\; W^+$ LO $\;$
 & $\;W^-$ NLO $\;$ & $\;W^-$ ME+PS$\;$ & $\; W^-$ LO $\;$ \\
\hline
\hline
$\; A_0\; $  & $ 0.399$  & $ 0.375 $ & $ 0.395$ 
             & $ 0.386 $ & $ 0.358 $ & $ 0.380$ \\
\hline
$A_1$  & $\hn -0.116 $ & $\hn -0.106 $ & $\hn -0.134 $
            & $\hn -0.109 $ & $ \hn -0.107 $ & $ \hn -0.130 $  \\
\hline
$A_2$  & $  0.318$  & $ 0.337 $ & $ 0.395 $ 
           & $ 0.310 $ & $ 0.327 $ & $ 0.379 $  \\
\hline 
$A_3$  & $ \hn -0.013$  & $ \hn -0.055 $ & $ \hn -0.014 $ 
           & $\hn -0.001 $ & $ 0.031 $ & $ \hn -0.001 $  \\
\hline 
$A_4$  & $ \hn -0.616$  & $\hn -0.565 $ & $ \hn -0.619 $ 
           & $ 0.497 $ & $  0.443 $ & $ 0.471 $  \\
\hline 
\end{tabular} 
\caption{The $A_i$ coefficients for $W$ production with $p_T^W>50$~GeV
and no restrictions on either the $W$ rapidity or the number of
associated jets. }
\label{AiTable} 
\end{table*}

Our results for the $A_i$ asymmetry coefficients of
\eqn{AiSDecomposition} are shown in \tab{AiTable}. Again taking the
difference between the NLO and ME+PS results as an estimate of the
theoretical uncertainty, we see that 
the uncertainty is under 10\%, except for $A_4$ in which it
is about 10\%, and $A_3$, which is very small but has a large
percentage shift between NLO and ME+PS. 
As mentioned previously, the $A_5$, $A_6$ and $A_7$
coefficients vanish at LO and for ME+PS.  At NLO, they
are much smaller than the current experimental and theoretical
uncertainties, and can therefore be neglected.

\begin{table*}
\vskip .4 cm
\begin{tabular}{||c||c| |c|c|c||}
\hline
 & CMS & NLO & ME+PS & LO \\
\hline
$W^+\;(f_L - f_R)$ &$\; 0.300 \pm 0.031 \pm 0.034\; $
 & $\; 0.308 \;$ & $\; 0.283\;$ & $\; 0.309\;$ \\
\hline
$W^-\;(f_L - f_R)$ & $0.226 \pm 0.031  \pm 0.050 $ 
  & $ 0.248$ & $ 0.222 $ & $ 0.235 $ \\
\hline
$W^+\;f_0$ &$\; 0.192 \pm 0.075 \pm 0.089\; $
  & $0.200$ & $0.187$ &  $0.198$ \\
\hline
$W^-\;f_0$ & $0.162 \pm 0.078  \pm 0.136 $ 
  & $0.193$ & $0.179$ & $ 0.190$ \\
\hline 
\end{tabular} 
\caption{A comparison of theoretical predictions for $f_L - f_R$ and $f_0$
  to preliminary CMS results~\cite{CMSMeasurement}. The first
  uncertainty in the CMS measurement is statistical and the second is
  systematic.
\label{PolarizationFractionCMSTable} 
}
\end{table*}

CMS recently presented a measurement of the polarization
fractions~\cite{CMSMeasurement}.  In
\tab{PolarizationFractionCMSTable} we compare our theoretical
predictions for $(f_L - f_R)$ and $f_0$ to the experimental ones.  Various
corrections for effects such as acceptance cuts have been applied by
CMS in order to produce the numbers in the table.  The NLO or ME+PS
predictions are both in excellent agreement with the data, within the
experimental uncertainties.  Large increases in the LHC data sets are
anticipated in the near future.  The improvement in experimental
precision that can be expected with these data should provide even
more incisive tests, possibly differentiating between the NLO and
ME+PS predictions.

\begin{figure*}[tbh]
\begin{minipage}[b]{1.\linewidth}
\includegraphics[clip,scale=.4]{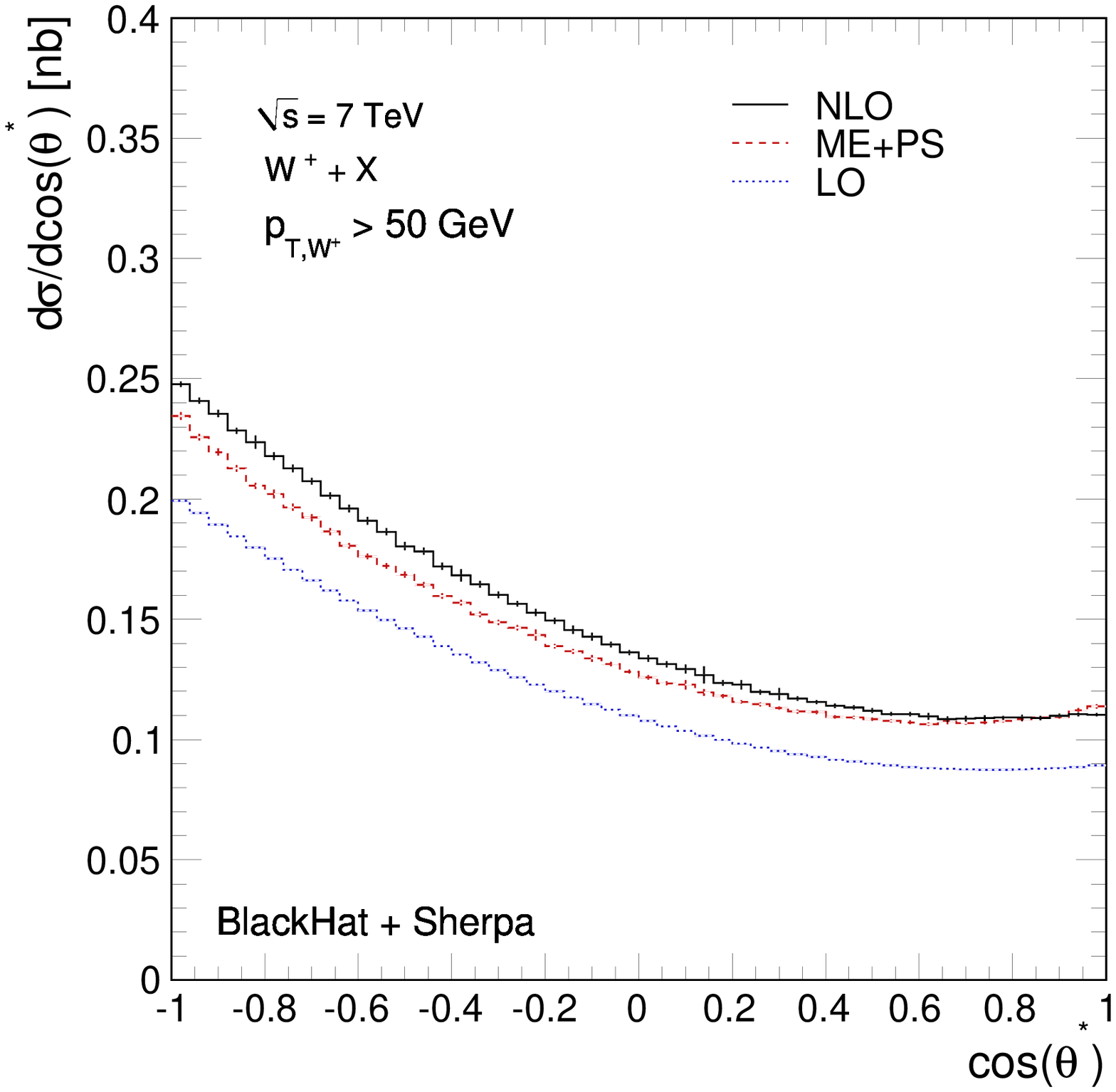}
\includegraphics[clip,scale=.4 ]{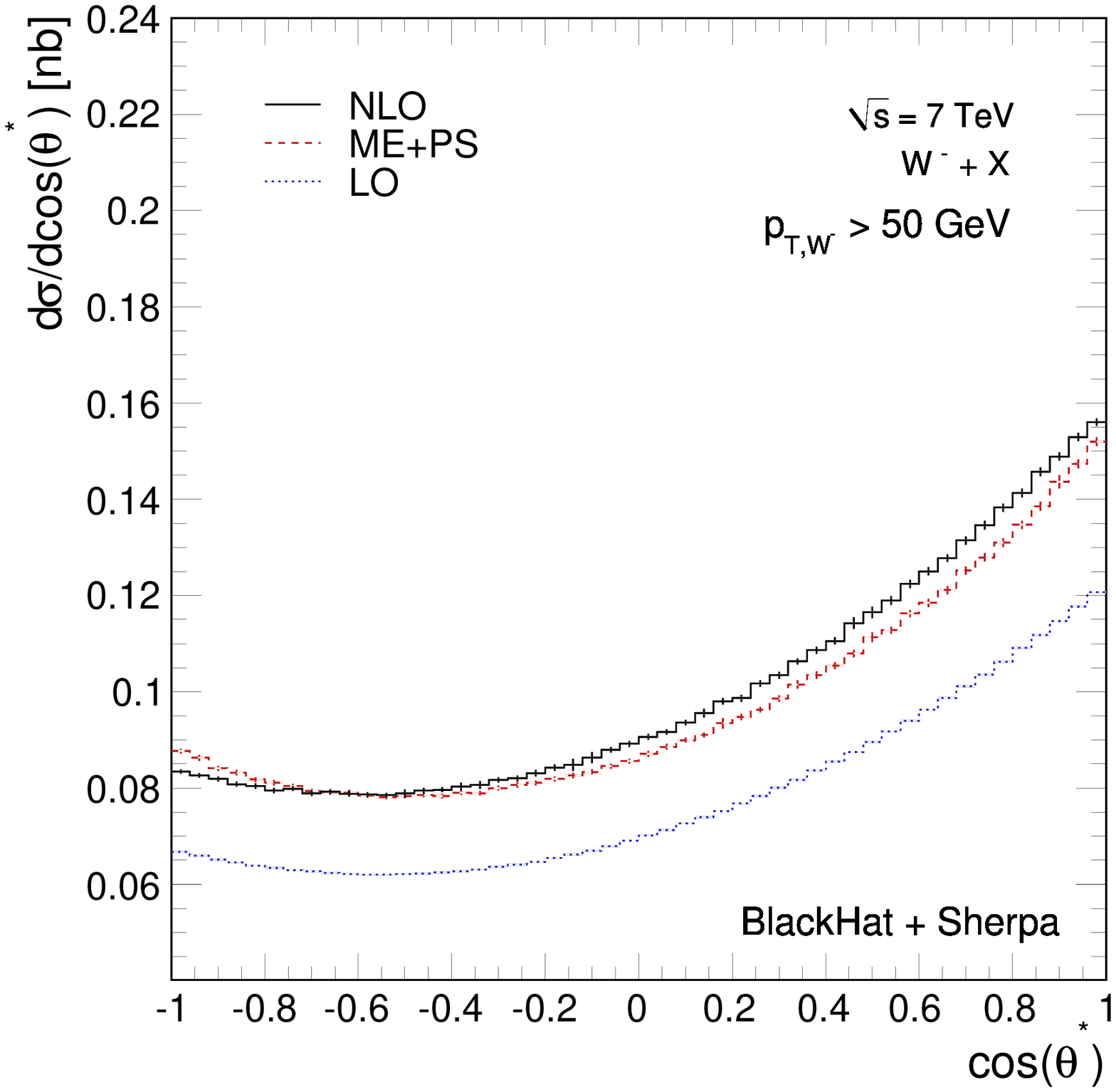}
\end{minipage}
\caption{The $\cos\ths$ distribution of the charged leptons for
  $W^\pm$ production with $p_T^W> 50$~GeV.
  The left plot shows the distribution in $W^+$ production,
  and the right plot, that in $W^-$ production. 
   Three different results are shown:
    the fixed-order NLO result represented by the solid (black)
  line; the ME+PS result represented by the dashed (red) line; and the
  fixed-order LO result represented by the dotted (blue)
  line. The thin vertical lines indicate the integration
  errors.}
\label{CosTheta_Figure}
\end{figure*}

\Fig{CosTheta_Figure} shows the $\cos \ths$ distributions for both
$W^+$ and $W^-$ bosons with $p_T^W > 50$~GeV.
These plots show that in the $W^+$ case, the charged anti-lepton prefers
to go backward with respect to the $W$ flight direction, while in the
$W^-$ case the charged lepton tends to go forward, in accordance with
the left-handed polarizations given in \tab{PolarizationFractionTable}.

\begin{figure*}[tbh]
\begin{minipage}[b]{1.\linewidth}
\includegraphics[clip,scale=.4]{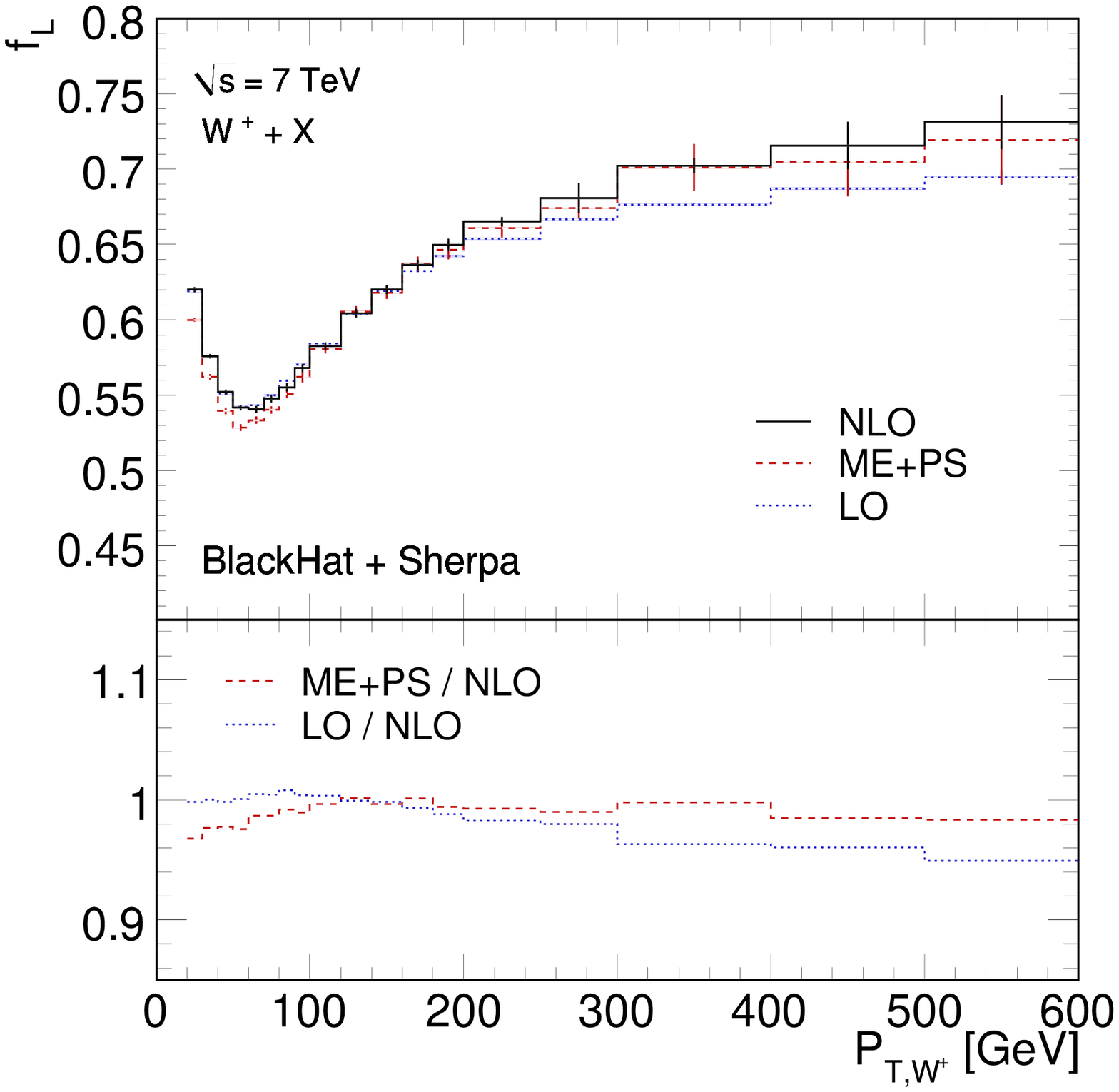}
\includegraphics[clip,scale=.4]{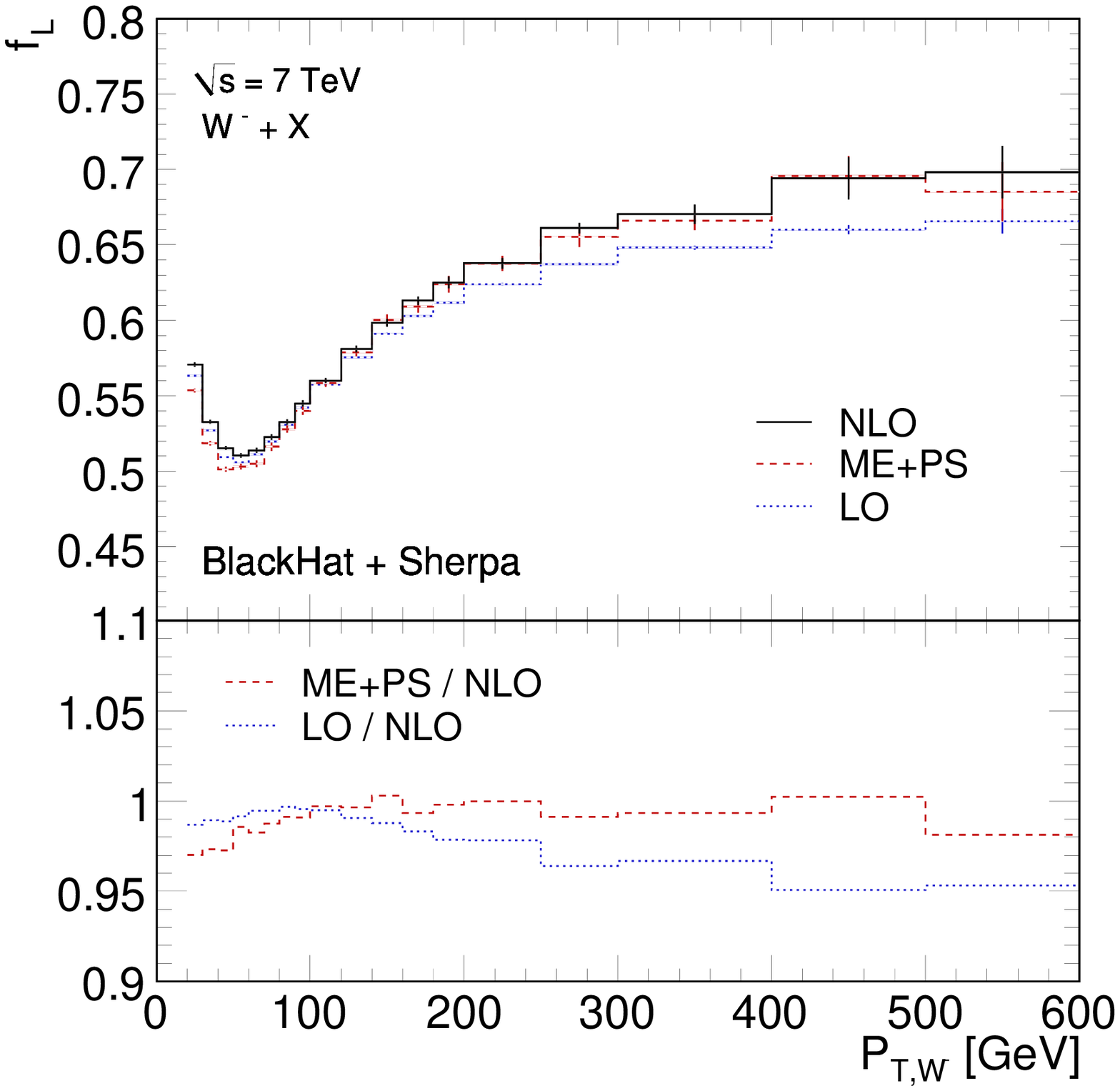} 
\end{minipage}
\caption{The left-handed polarization fraction $f_L$ as a function of
  $p_T^W$ for $W^\pm$ production at the LHC. The left panel gives
  the $W^+$ case and the right panel the $W^-$ case.  Three different
  results are shown: the fixed-order NLO result represented by the
  solid (black) line; the ME+PS result represented by the dashed (red)
  line; and the fixed-order LO result represented by the dotted
  (blue) line.  The thin vertical lines indicate the integration
  errors.  The lower panels show ratios normalized to the NLO 
  result.
\label{FLFigure}}
\end{figure*}

\begin{figure*}[tbh]
\begin{minipage}[b]{1.\linewidth}
\includegraphics[clip,scale=.4]{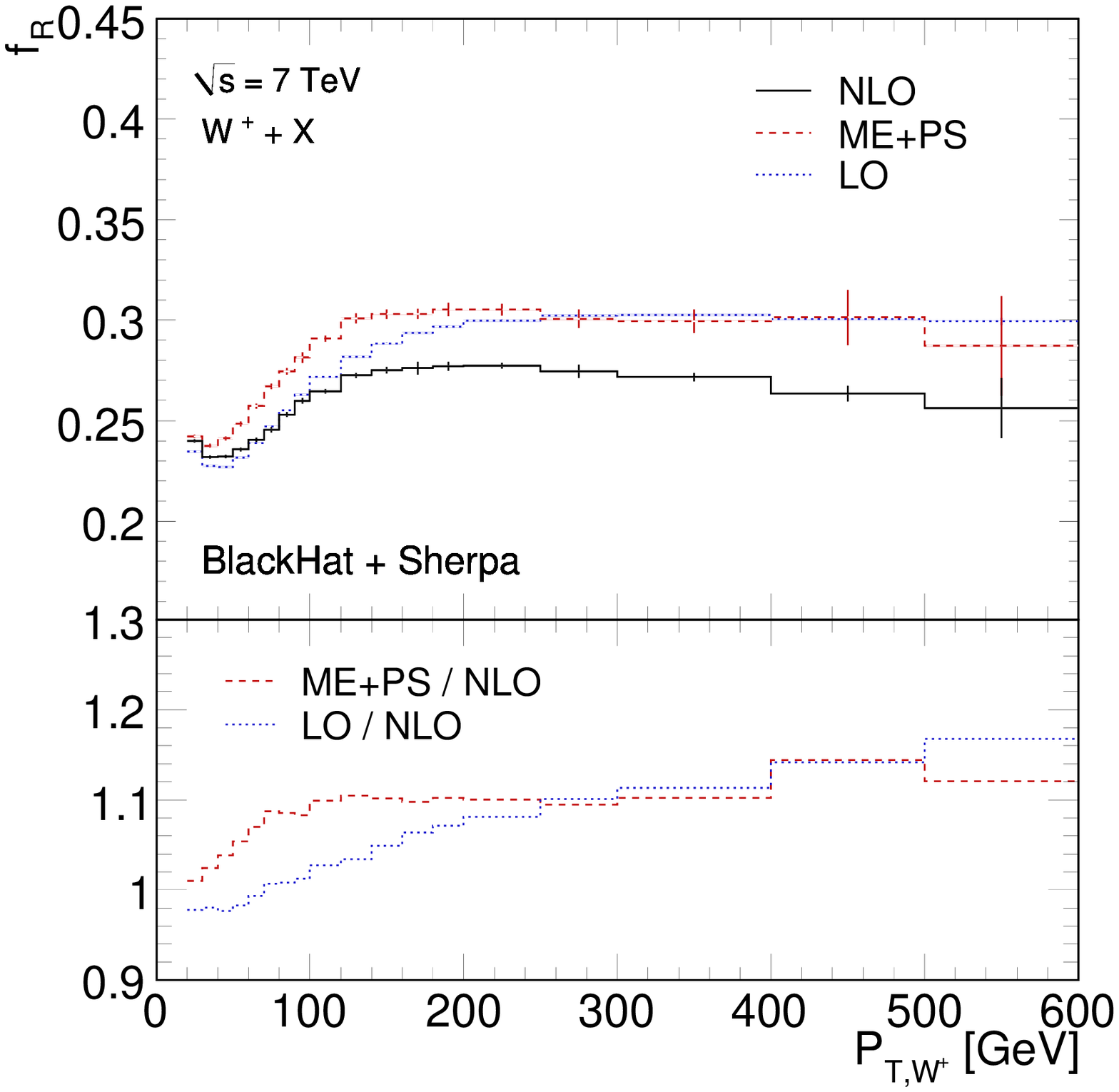} 
\includegraphics[clip,scale=.4]{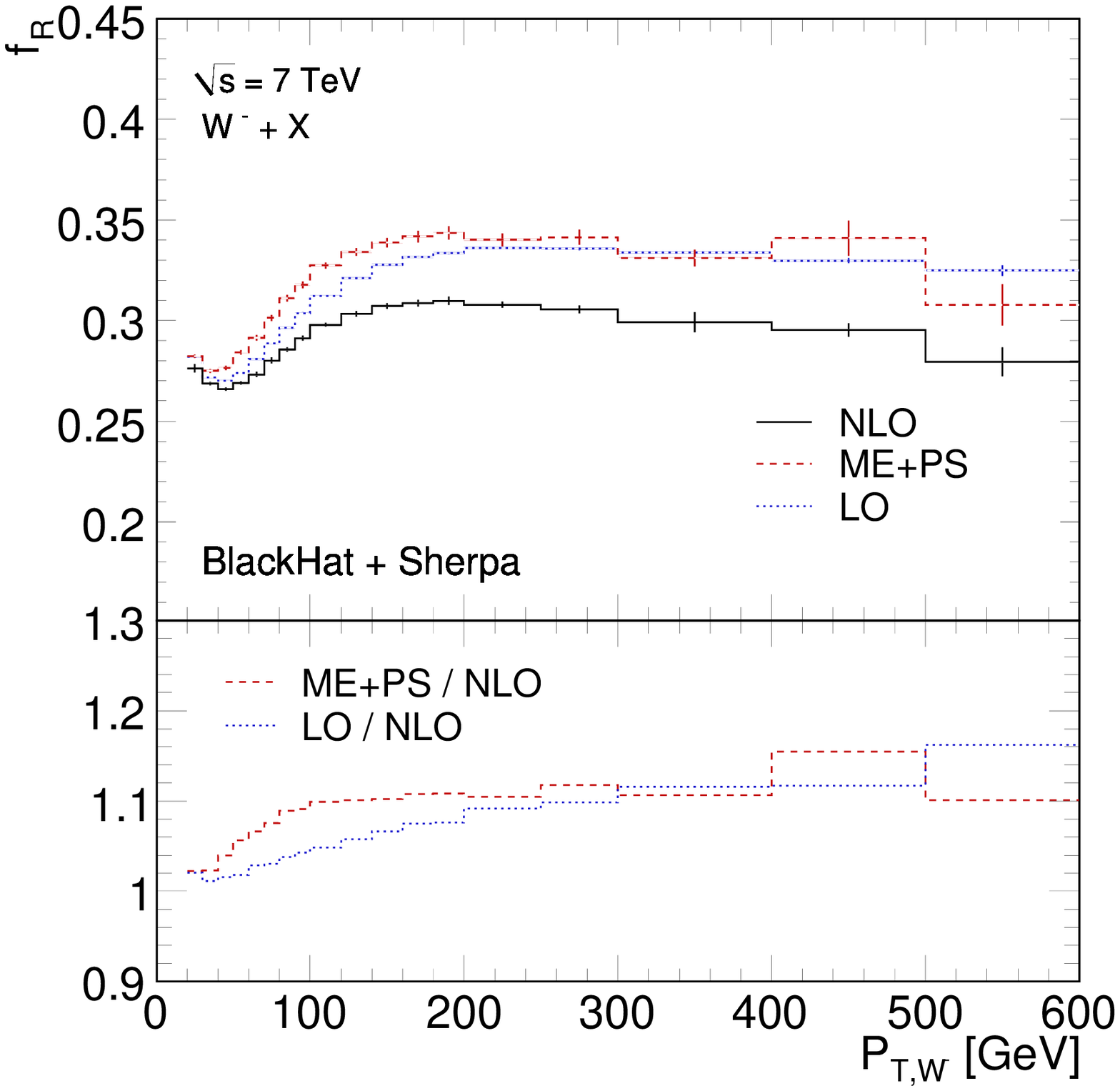}
\end{minipage}
\caption{The coefficient $f_R$ as a function of $p_T^W$.
The left panel is for $W^+$ and the right panel for $W^-$.
The format is the same as in \fig{FLFigure}.
}
\label{FRFigure}
\end{figure*}

\begin{figure*}[tbh]
\begin{minipage}[b]{1.\linewidth}
\includegraphics[clip,scale=.4]{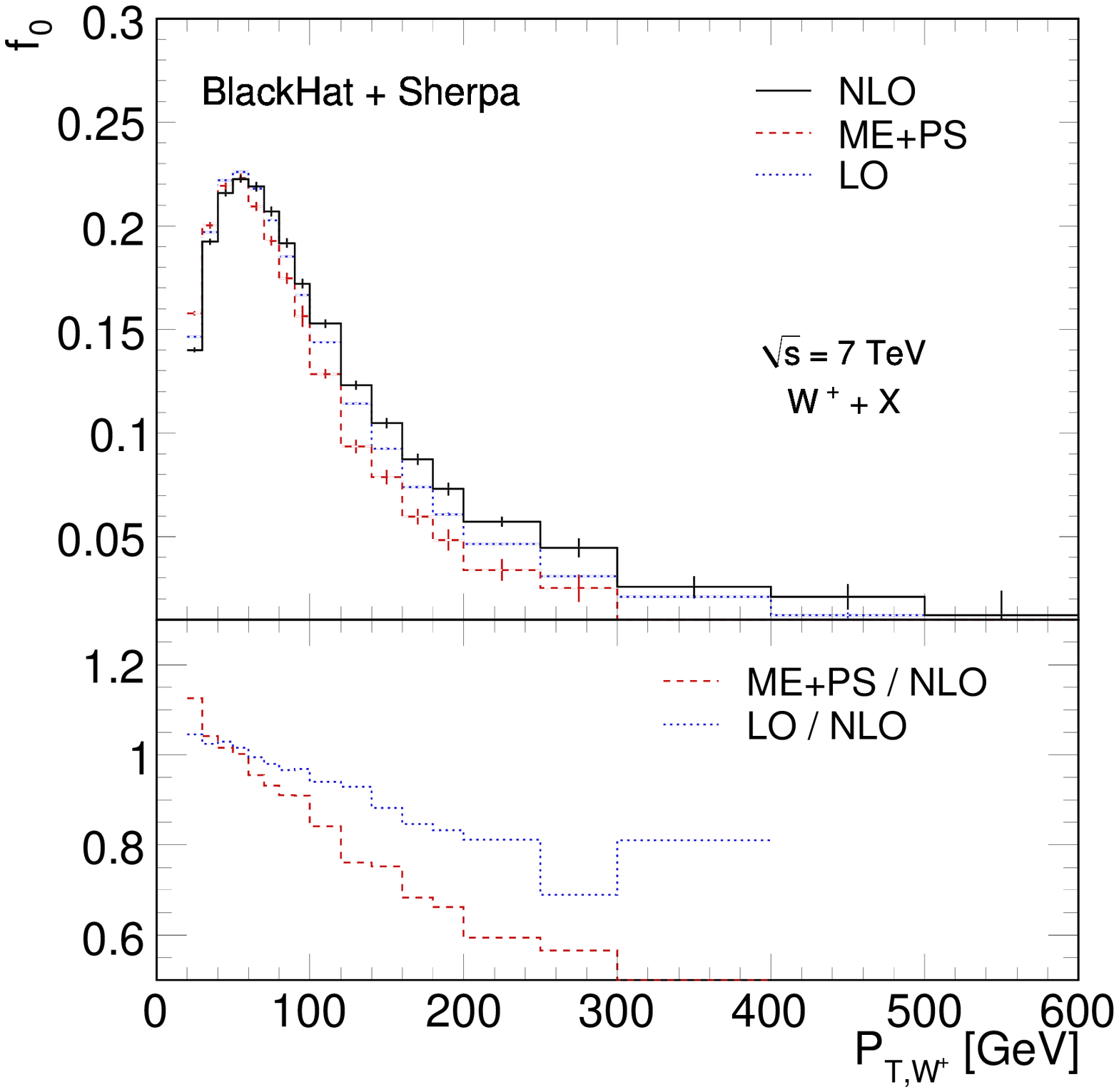}
\includegraphics[clip,scale=.4]{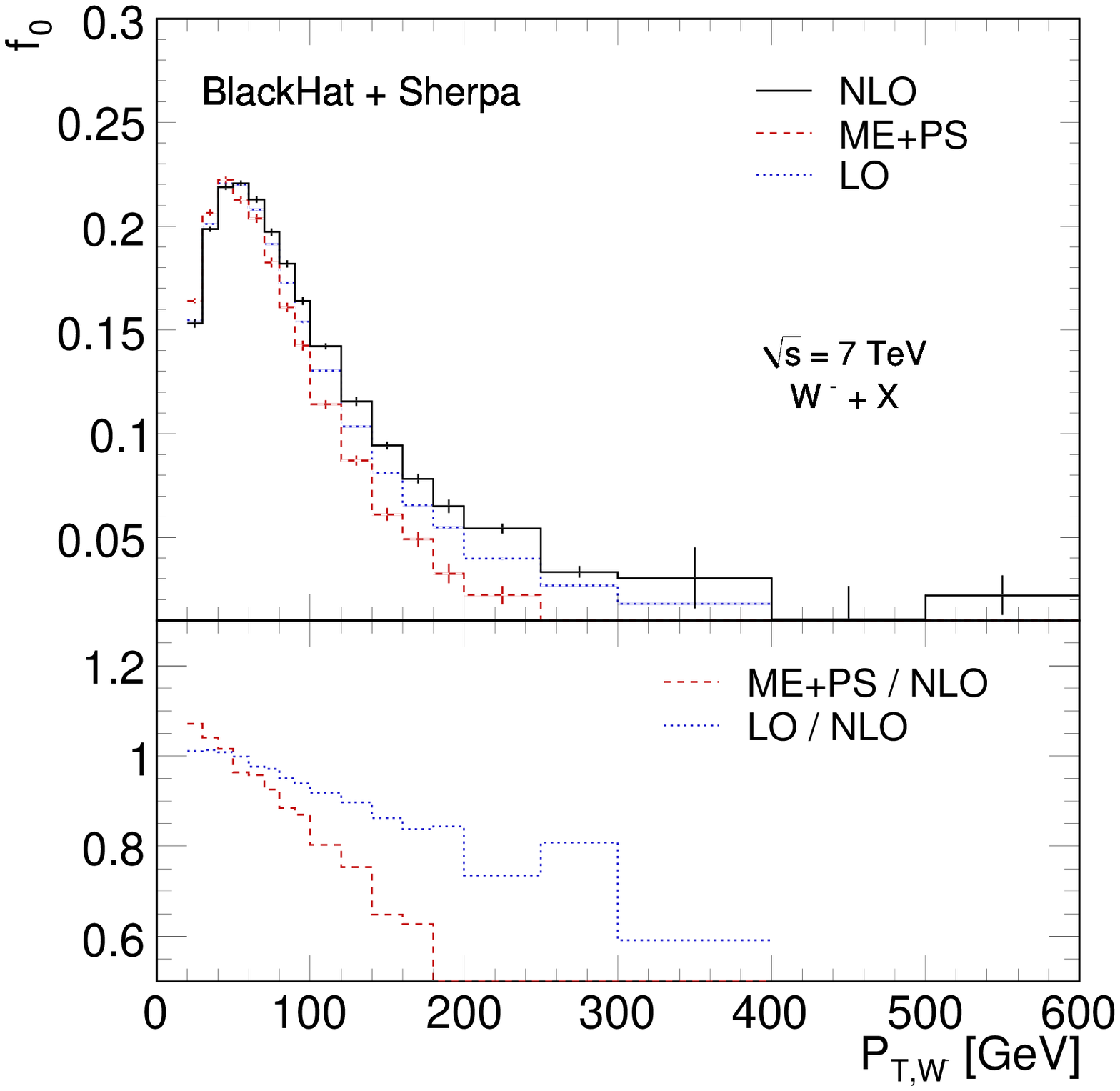}
\end{minipage}
\caption{The coefficient $f_0 = A_0/2$ as a function of $p_T^W$.  It vanishes
  at large $p_T^W$ by the equivalence theorem.  The left panel is for $W^+$
  and the right panel for $W^-$. The format is the same as in \fig{FLFigure}.
 }
\label{F0Figure}
\end{figure*}

\Fig{FLFigure} displays the polarization fraction $f_L$
for both $W^+$ and $W^-$ bosons at the LHC, as a function of the
vector bosons'
transverse momenta.  There are again no rapidity cuts and no explicit
jet requirements (in contrast to the setup for
\figs{WpComparisonFigure}{ZComparisonFigure}).
The fraction $f_L$ climbs to around $0.7$
at high $p_T^W$.  The NLO predictions are a bit higher than the LO
and ME+PS ones.  \Fig{FRFigure} contains the corresponding plots
for the right-handed fraction $f_R$.  Although this
component rises initially, by 150~GeV it stabilizes
between $0.25$ to $0.30$ for the $W^+$ case. For the $W^-$ case,
it is a bit higher. For $f_R$ the NLO predictions are a bit lower
than the LO and ME+PS ones.  In any case, to compensate for these
rises, the longitudinal component $f_0$ falls rapidly with increasing $p_T^W$,
as illustrated in \fig{F0Figure}.  As mentioned previously, the
decline of $f_0$ is due to the equivalence theorem.  A
rather striking feature of these plots is how small the difference is
between the $W^-$ and $W^+$ cases, showing that the effect is essentially
the same for both signs.

\begin{figure*}[tbh]
\begin{minipage}[b]{1.\linewidth}
\includegraphics[clip,scale=.5]{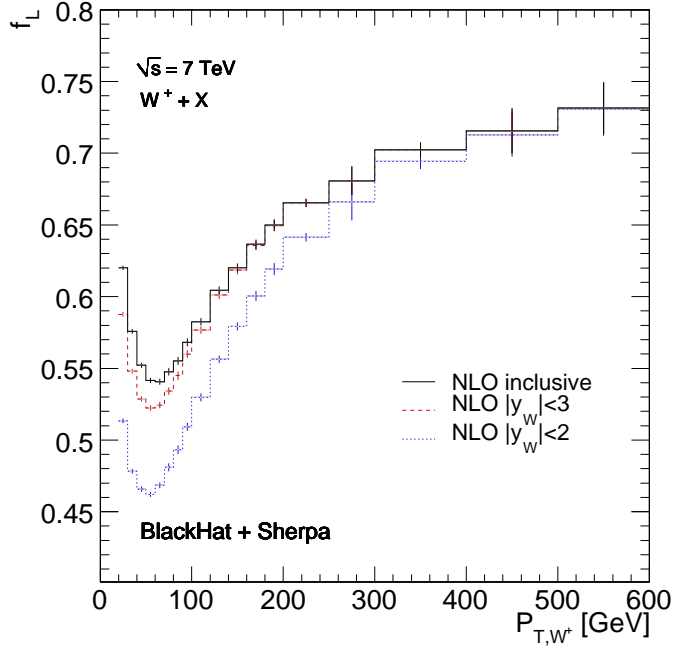}
\end{minipage}
\caption{The $f_L$ polarization fraction in NLO QCD for the 
$W^+$ case. The solid (black) curve
is with no with no rapidity cut, the dashed (red) curve with a cut 
$|y_{W}|< 2$, and the dotted (blue) curve with $|y_{W}| <3$.  (The vertical
axis starts at $f_L = 0.4$ to visually separate the curves.)
}
\label{FL_ycutFigure}
\end{figure*}

\begin{figure*}[tbh]
\begin{minipage}[b]{1.\linewidth}
\includegraphics[clip,scale=.4]{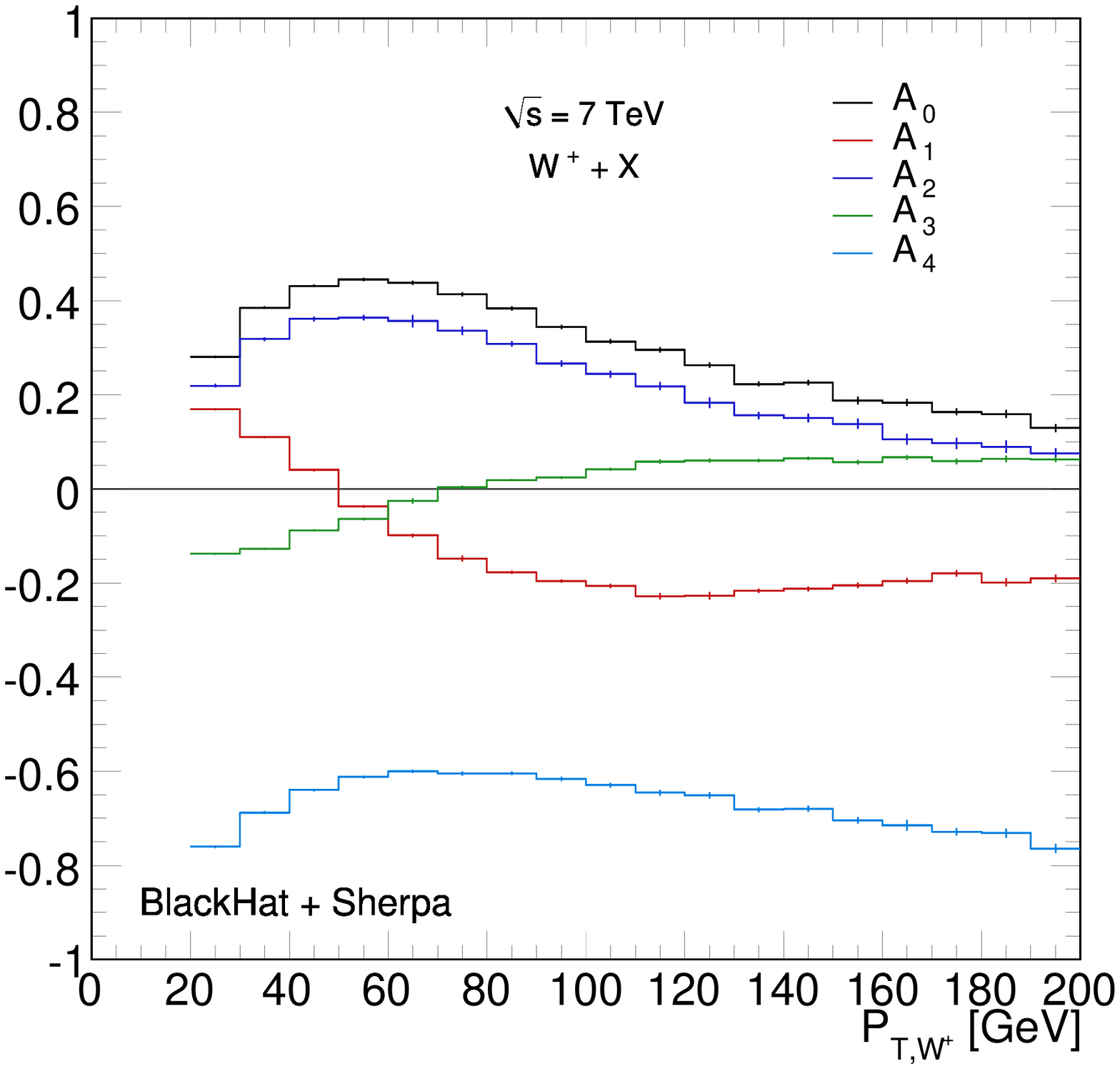}
\includegraphics[clip,scale=.4]{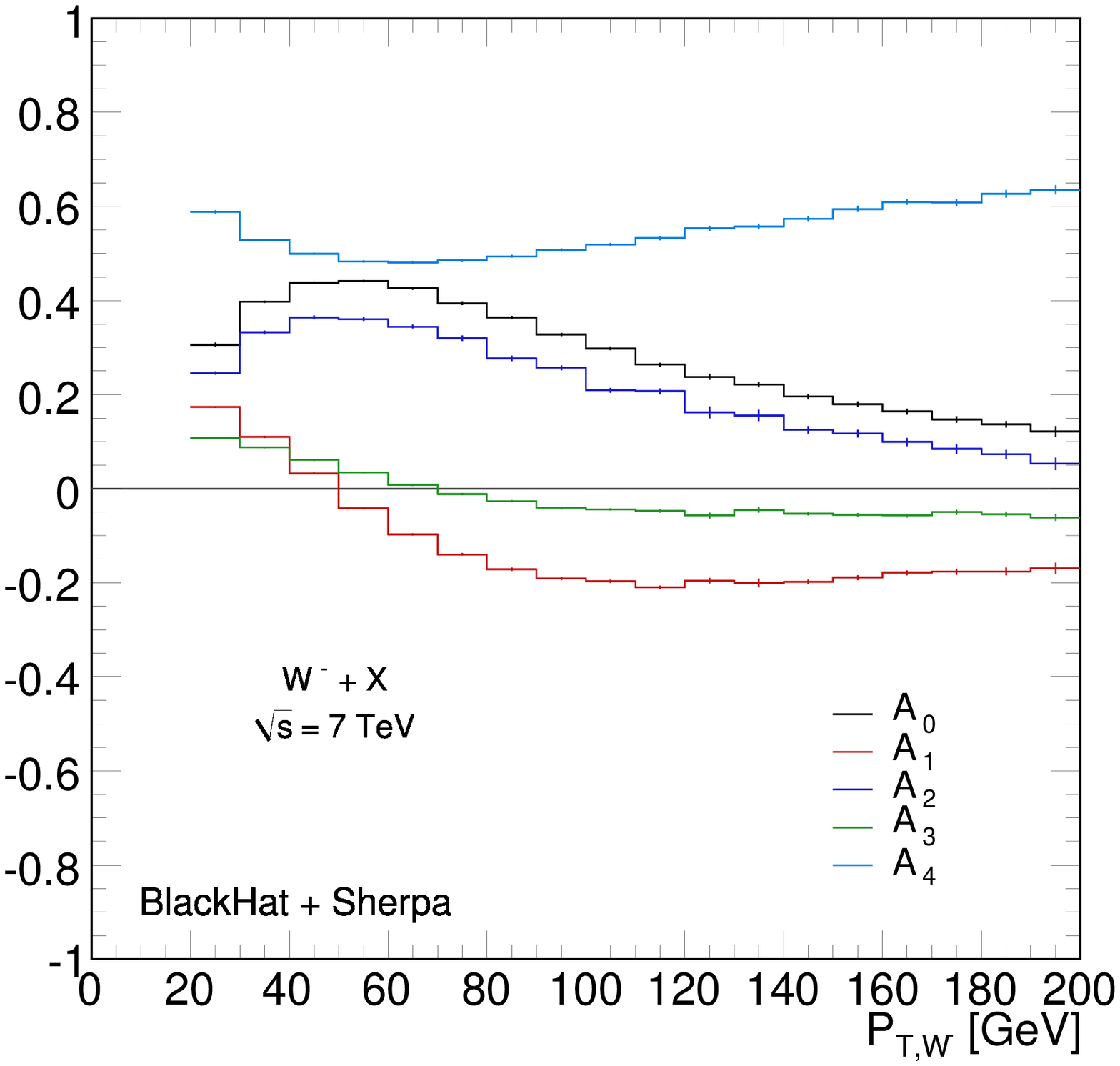}
\end{minipage}
\caption{The first five $A_i$ as a function of $p_T^W$, at NLO.
The left panel is for $W^+$ production, and the right panel for $W^-$
production. 
Starting from the top and proceeding downwards,
the left-hand sides of the curves at low $p_T^W$ are in the order
$A_0, A_2, A_1, A_3, A_4$ for the left panel.
  For the right panel they are in the order
$A_4,A_0,A_2,A_1,A_3$. 
}
\label{AiFigure}
\end{figure*}

Another interesting plot is the left-handed polarization fraction
$f_L$ as a function of $p_T^W$, but with different rapidity cuts
imposed on the vector-boson $y_{W}$.  In \fig{FL_ycutFigure}
we show the three curves, one with no rapidity cut, 
one with $|y_{W}|< 2$,
and one with $|y_{W}| <3$.  At low $p_T^W$, the polarization
fraction in the central region with $|y_W| < 2$ is lower than for
$|y_W| < 3$, which in turn is lower than the fraction
with no rapidity cut imposed.  At low $p_T$, $f_L$
picks up a large left-handed polarization in the forward and backward
regions from the beam-axis effect described in
\sect{DynamicsSection}, while the transverse effect has not fully kicked
in.  By a transverse momentum of 150 GeV, the effect of the 
$|y_W|<3$ cut has essentially disappeared; and by
a transverse momentum of 350 GeV, the effect of any rapidity cut
has essentially disappeared.  This demonstrates that the large
polarization at high $p_T^W$ comes from central rapidities.
One could impose a rapidity cut on the $W$ boson in order to separate
the beam-axis polarization effect from the transverse one at
all vector-boson transverse momenta.

Finally in \fig{AiFigure}, we compare the five $A_i$ coefficients for
the cases of $W^+$ and $W^-$ using NLO QCD.  Up to sign flips for $A_3$
and $A_4$  we see little difference between the two cases,
as a function of $p_T^W$.   The approximate equality $A_0 \approx A_2$ is
due to the (frame-independent) Lam-Tung relation~\cite{LamTung}, which
holds at LO, but is violated at NLO.


\section{Conclusions}
\label{ConclusionSection}

Prompt $W$ vector bosons of both signs, when produced at moderate
to high transverse momentum at the LHC, are predominantly polarized
left-handedly~\cite{W3jDistributions,Polarization}.
In this paper, we presented a detailed study of this phenomenon
and of its underyling mechanism.
The effect, which superficially appears to violate CP, 
actually arises from a combination of the left-handed nature of
the electroweak charged-current interaction,
the prevalence of valence quarks in the $pp$ initial state (which is
not charge-conjugation invariant), and properties of the 
short-distance matrix elements.

We found that a simple estimate, assuming $90^\circ$ scattering 
in the partonic center-of-mass for the dominant process $qg\to Wq'$,
leads to 80\% left-handed polarization for $W$ bosons at large $p_T^W$.
This high degree of polarization is somewhat reduced by kinematic
effects in this subprocess, as well as dilution from other
subprocesses, especially at lower $p_T^W$.
Nevertheless, with a cut of $p_T^W> 50$~GeV we find
that most $W$ bosons are left-handed, and the remainder are split between
right-handed and longitudinal states. As $p_T^W$ increases, the
left-handed polarization fraction can rise as high as 70\%.
This is remarkably close to the simple upper estimate of 80\%.  The
$Z$ boson polarization is similar, though a bit smaller at around 60\%,
due to dilution from right-handed $u$ quarks in the initial state.
The effect in $Z$ decays will be harder to see experimentally, 
however, as the analyzing power in decays to charged leptons
is only about 15\% (versus 100\% for $W\to l\nu$).

We studied the polarization dominance using LO, NLO and 
ME+PS QCD calculations.  The effect is
theoretically robust, and higher-order QCD
corrections are small.  The difference between NLO and ME+PS
predictions suggests a 10\% theoretical uncertainty in
the polarization fractions.  It should be possible in the future to
improve the predictions, when there are no tagged jets, or even one
tagged jet, in at least two ways.  First, it should be possible to
generate \Wj{}-jet events at NLO accuracy incorporating a parton
shower, as a program already exists for the closely-related case of
\Zj{}-jet production~\cite{ZjetPOWHEG} based on the {\sc POWHEG}
method~\cite{POWHEG}. 
The {\sc MC@NLO} approach~\cite{MCNLO} should also
be feasible.
Second, the computation of \Wj{}-jet production at NNLO
in QCD may become feasible before long~\cite{NNLOWj}, making it possible
to study the polarization observables at one higher order
in $\alpha_s$.

We provided numerical tables and plots for the polarization
fractions and compared to the recent preliminary measurement
by the CMS collaboration~\cite{CMSMeasurement}.
The theoretical predictions are in excellent agreement with this initial 
measurement, given its relatively sizable uncertainties.  
More detailed comparisons will be possible with
increased data sets from the ongoing run of the LHC.

It may also be possible to use the $W$ polarization phenomenon as a
probe of polarized gluon distribution functions, if a sufficient number
of $W$ bosons can be produced at moderate transverse momentum in
polarized proton collisions.  Asymmetries in
inclusive (low $p_T^W$) $W$ boson production in polarized $pp$ collisions
have recently been studied at RHIC~\cite{STARPHENIX}.

The large left-handed polarization of prompt vector bosons at
high-transverse momentum is a robust theoretical prediction, stable
against both QCD corrections and the emission of additional jets.
It leads to very different distributions for the positively and 
negatively charged
decay leptons, as well as for the neutrinos.  The situation for
top-quark pair production is very different.  The initial state is
predominantly all-gluon, and hence CP-invariant, so no such asymmetry
is possible.  While top quark decay produces $W^+$ bosons that are
about 70\% longitudinal, 30\% left-handed, the $W^-$ bosons from
$\bar{t}$ decay are 70\% longitudinal, 30\% right-handed, so in this
case the positively and negatively charged leptons have very similar
distributions.  The decay to $WW$ pairs of a heavy Higgs boson 
(sufficiently heavy to decay to moderately high-$p_T$ $W$ bosons)
is another signal process which should not display single-$W$
asymmetries, because of the
spin-0 nature of the Higgs boson.  The same will hold true for $W$
bosons arising from many sources beyond Standard-Model physics.  These
distinctions give $W$ polarization the potential to be a powerful
discriminant for interesting Standard-Model signals and new physics
beyond the Standard Model.

{\bf Note added:} After the first version of this article appeared,
CMS~\cite{CMSWpolpaper} has produced a paper on the $W$ polarization
measurement at the LHC; also, CDF~\cite{CDFZpol} has measured several
$A_i$ coefficients for $Z$ production with transverse momentum at the
Tevatron, using the Collins-Soper frame.

\section*{Acknowledgments}

We thank Carola Berger for contributions at the initial
stages of this project. We are grateful to Oliver Buchm\"{u}ller, Jad
Marrouche, Roberto Peccei, Michael Peskin, Jeff Richman, David Saltzberg, Paris
Sphicas and Markus Stoye for useful conversations. This
research was supported by the US Department of Energy under contracts
DE--FG03--91ER40662 and DE--AC02--76SF00515.  DAK's research is
supported by the European Research Council under Advanced Investigator
Grant ERC--AdG--228301.  HI's work is supported by a grant from the US
LHC Theory Initiative through NSF contract PHY--0705682.  This
research used resources of Academic Technology Services at UCLA,
PhenoGrid using the GridPP infrastructure.


\begin{thebibliography}{99}

\bibitem{W3jDistributions}
C.~F.~Berger {\it et al.},
Phys.\ Rev.\  D {\bf 80}, 074036 (2009)
[0907.1984 [hep-ph]].


\bibitem{ESW}
R.~K.~Ellis, W.~J.~Stirling and B.~R. Webber,
{\it QCD and Collider Physics} (Cambridge University Press, 1996).


\bibitem{LOPrograms}
T.~Stelzer and W.~F.~Long,
Comput.\ Phys.\ Commun.\  {\bf 81}, 357 (1994)
[hep-ph/9401258];\\
%
A.~Pukhov {\it et al.},
hep-ph/9908288;\\
%
M.~L.~Mangano, M.~Moretti, F.~Piccinini, R.~Pittau and A.~D.~Polosa,
JHEP {\bf 0307}, 001 (2003)
[hep-ph/0206293].

\bibitem{Amegic}
F.~Krauss, R.~Kuhn and G.~Soff,
JHEP {\bf 0202}, 044 (2002)
[hep-ph/0109036];\\
%
T.~Gleisberg and F.~Krauss,
Eur.\ Phys.\ J.\  C {\bf 53}, 501 (2008)
[0709.2881 [hep-ph]].

\bibitem{Matching}
S.~Catani, F.~Krauss, R.~Kuhn and B.~R.~Webber,
JHEP {\bf 0111}, 063 (2001)
[hep-ph/0109231];\\
M.~Mangano,
presented at the Fermilab ME/MC TuningWorkshop, October 4, 2004.

\bibitem{BHPRL}
C.~F.~Berger {\it et al.},
Phys.\ Rev.\ Lett.\ {\bf 102}, 222001 (2009)
[0902.2760 [hep-ph]].

\bibitem{Ellis3j}
R.~K.~Ellis, K.~Melnikov and G.~Zanderighi,
Phys.\ Rev.\  D {\bf 80}, 094002 (2009)
[0906.1445 [hep-ph]];\\
%
K.~Melnikov and G.~Zanderighi,
Phys.\ Rev.\  D {\bf 81}, 074025 (2010)
[0910.3671 [hep-ph]].

\bibitem{BHZ3j}
C.~F.~Berger {\it et al.},
Phys.\ Rev.\  D {\bf 82}, 074002 (2010)
[1004.1659 [hep-ph]].

\bibitem{BHW4j}
C.~F.~Berger {\it et al.},
Phys.\ Rev.\ Lett.\ {\bf 106}, 092001 (2011)
[1009.2338 [hep-ph]]. 

\bibitem{Polarization}
C.~F.~Berger {\it et al.},
PoS {\bf RADCOR2009}, 002 (2009)
[0912.4927 [hep-ph]].

\bibitem{Equivalence}
J.~M.~Cornwall, D.~N.~Levin and G.~Tiktopoulos,
Phys.\ Rev.\  D {\bf 10}, 1145 (1974)
[Erratum-ibid.\  D {\bf 11}, 972 (1975)].

\bibitem{CMSMeasurement}
P.~Harris [CMS preliminary], presented at
Rencontres de Moriond EW 2011 (March 13-20, 2011).

\bibitem{CollinsSoper}
J.~C.~Collins and D.~E.~Soper,
Phys.\ Rev.\  D {\bf 16} (1977) 2219.

\bibitem{LamTung}
C.~S.~Lam and W.-K.~Tung,
Phys.\ Rev.\  D {\bf 21}, 2712 (1980).

\bibitem{Hagiwara1}
K.~Hagiwara, K.~Hikasa and N.~Kai,
Phys.\ Rev.\ Lett.\  {\bf 52}, 1076 (1984).

\bibitem{Mirkes}
E.~Mirkes,
Nucl.\ Phys.\  B {\bf 387}, 3 (1992).

\bibitem{MirkesOhnemus}
E.~Mirkes and J.~Ohnemus,
Phys.\ Rev.\  D {\bf 50}, 5692 (1994)
[hep-ph/9406381];\\
%
E.~Mirkes and J.~Ohnemus,
Phys.\ Rev.\  D {\bf 51}, 4891 (1995)
[hep-ph/9412289].

\bibitem{Hagiwara2}
K.~Hagiwara, K.~Hikasa and H.~Yokoya,
Phys.\ Rev.\ Lett.\  {\bf 97}, 221802 (2006)
[hep-ph/0604208].

\bibitem{CDFRunI}
D.~E.~Acosta {\it et al.}  [CDF Collaboration],
Phys.\ Rev.\  D {\bf 73}, 052002 (2006)
[hep-ex/0504020].

\bibitem{BlackHatI}
C.~F.~Berger {\it et al.},
Phys.\ Rev.\ D {\bf 78}, 036003 (2008)
[0803.4180 [hep-ph]].

\bibitem{Sherpa}
T.~Gleisberg {\it et al.}, 
JHEP {\bf 0402}, 056 (2004)
[hep-ph/0311263];\\
%
T.~Gleisberg {\it et al.},
JHEP {\bf 0902}, 007 (2009)
[0811.4622 [hep-ph]].

\bibitem{CTEQ6M}
J.~Pumplin {\it et al.},
JHEP {\bf 0207}, 012 (2002)
[hep-ph/0201195].

\bibitem{KomStirling}
C.~H.~Kom and W.~J.~Stirling,
Eur.\ Phys.\ J.\  C {\bf 69}, 67 (2010)
[1004.3404 [hep-ph]];
%
Eur.\ Phys.\ J.\  C {\bf 71}, 1546 (2011)
[1010.2988 [hep-ph]].

\bibitem{SISCONE}
G.~P.~Salam and G.~Soyez,
JHEP {\bf 0705}, 086 (2007)
[0704.0292 [hep-ph]].

\bibitem{MEPS}
S.~H{\"o}che, F.~Krauss, S.~Schumann and F.~Siegert,
JHEP {\bf 0905}, 053 (2009)
[0903.1219 [hep-ph]].

\bibitem{CSShower}
S.~Schumann and F.~Krauss,
JHEP {\bf 0803}, 038 (2008)
[0709.1027 [hep-ph]];\\
%
S.~H{\"o}che, S.~Schumann, F.~Siegert,
Phys.\ Rev.\  D {\bf 81}, 034026 (2010)
[0912.3501 [hep-ph]].

\bibitem{CS}
S.~Catani and M.~H.~Seymour,
Nucl.\ Phys.\  B {\bf 485}, 291 (1997)
[Erratum-ibid.\  B {\bf 510}, 503 (1998)]
[hep-ph/9605323].

\bibitem{MSTW}
A.~D.~Martin, W.~J.~Stirling, R.~S.~Thorne and G.~Watt,
Eur.\ Phys.\ J.\  C {\bf 63}, 189 (2009)
[0901.0002 [hep-ph]].

\bibitem{Z4Partons}
Z.~Bern, L.~J.~Dixon and D.~A.~Kosower,
Nucl.\ Phys.\  B {\bf 513}, 3 (1998)
[hep-ph/9708239].

\bibitem{AntennaIntegrator}
A.~van Hameren and C.~G.~Papadopoulos,
Eur.\ Phys.\ J.\  C {\bf 25}, 563 (2002)
[hep-ph/0204055];\\
%
T.~Gleisberg, S.~H\"oche and F.~Krauss,
0808.3672 [hep-ph].

\bibitem{SherpaCode}
http://www.hepforge.org/sherpa/


\bibitem{ZjetPOWHEG}
S.~Alioli, P.~Nason, C.~Oleari and E.~Re,
JHEP {\bf 1101}, 095 (2011)
[1009.5594 [hep-ph]].

\bibitem{POWHEG}
S.~Frixione, P.~Nason and C.~Oleari,
JHEP {\bf 0711}, 070 (2007)
[0709.2092 [hep-ph]];\\
%
S.~Alioli, P.~Nason, C.~Oleari and E.~Re,
JHEP {\bf 0807}, 060 (2008)
[0805.4802 [hep-ph]].

\bibitem{MCNLO}
S.~Frixione and B.~R.~Webber,
JHEP {\bf 0206}, 029 (2002)
[hep-ph/0204244]; \\
%
S.~Frixione, P.~Nason and B.~R.~Webber,
JHEP {\bf 0308}, 007 (2003)
[hep-ph/0305252].

\bibitem{NNLOWj}
P.~Bolzoni, S.~Moch, G.~Somogyi and Z.~Tr\'ocs\'anyi,
JHEP {\bf 0908} (2009) 079
[0905.4390 [hep-ph]];\\
%
A.~Daleo, A.~Gehrmann-De Ridder, T.~Gehrmann and G.~Luisoni,
JHEP {\bf 1001} (2010) 118
[0912.0374 [hep-ph]];
PoS {\bf RADCOR2009} (2010) 062
[1001.2397 [hep-ph]];
PoS {\bf DIS2010} (2010) 122;\\
%
E.~W.~N.~Glover and J.~Pires,
JHEP {\bf 1006} (2010) 096
[1003.2824 [hep-ph]];
%
Nucl.\ Phys.\ Proc.\ Suppl.\  {\bf 205-206} (2010) 176
[1006.1849 [hep-ph]];\\
%
R.~Boughezal, A.~Gehrmann-De Ridder and M.~Ritzmann,
PoS {\bf DIS2010} (2010) 101;
%
JHEP {\bf 1102} (2011) 098
[1011.6631 [hep-ph]];\\
%
P.~Bolzoni, G.~Somogyi and Z.~Tr\'ocs\'anyi,
JHEP {\bf 1101} (2011) 059
[1011.1909 [hep-ph]].

\bibitem{STARPHENIX}
M.~M.~Aggarwal {\it et al.}  [STAR Collaboration],
Phys.\ Rev.\ Lett.\  {\bf 106}, 062002 (2011)
[1009.0326 [hep-ex]];\\
%
A.~Adare {\it et al.}  [PHENIX Collaboration],
Phys.\ Rev.\ Lett.\  {\bf 106}, 062001 (2011)
[1009.0505 [hep-ex]].

\bibitem{CMSWpolpaper}
S.~Chatrchyan {\it et al.}  [CMS Collaboration],
1104.3829 [hep-ex].

\bibitem{CDFZpol}
T.~Aaltonen {\it et al.}  [CDF Collaboration],
1103.5699 [hep-ex].

\end{thebibliography}
\end{document}